\documentclass{article}

\usepackage[preprint]{icml2026}

\usepackage[breakable]{tcolorbox}
\usepackage{hyperref}
\usepackage[utf8]{inputenc}
\usepackage{xspace}
\usepackage{graphicx}
\usepackage{amsmath}
\usepackage{algorithmic}
\usepackage{array}
\usepackage[caption=false,font=footnotesize,labelfont=sf,textfont=sf]{subfig}
\usepackage{booktabs}
\usepackage{url}
\usepackage{cleveref}
\usepackage{xcolor}         
\usepackage{xspace}
\usepackage{multirow}
\usepackage{colortbl}
\usepackage{array}
\usepackage{graphicx}
\usepackage{mdframed}
\usepackage{pifont}
\usepackage{amsmath}
\usepackage{amsmath}
\usepackage{listings}
\usepackage{makecell}

\newcommand{\subcap}[1]{\vspace{2pt}\par\centering\footnotesize #1\par}
\newcommand{\sys}{VulnLLM-R\xspace}

\newcommand{\base}{Qwen2.5-7B-Instruct\xspace}
\newcommand{\dsr}{DeepSeek-R1\xspace}
\newcommand{\qwq}{QwQ-32B\xspace}

\newcommand{\claude}{Claude-3.7-Sonnet\xspace}

\newcommand{\omini}{o3-mini\xspace}
\newcommand{\othree}{o3\xspace}
\newcommand{\gptfourone}{GPT-4.1\xspace}
\newcommand{\seccodeplt}{SecCodePLT\xspace}
\newcommand{\juliet}{Juliet 1.3\xspace}
\newcommand{\primevul}{PrimeVul\xspace}
\newcommand{\secllm}{SecLLMHolmes\xspace}
\newcommand{\arvo}{ARVO\xspace}
\newcommand{\sven}{SVEN\xspace}

\newcommand{\qweninstruct}{Qwen2.5-7B-Instruct\xspace}
\newcommand{\qwenthree}{Qwen3-8B\xspace}
\newcommand{\codeql}{CodeQL\xspace}
\newcommand{\aflpp}{AFL++\xspace}
\newcommand{\gtwofuzz}{G\textsuperscript{2}Fuzz\xspace}

\newcommand{\jazzer}{Jazzer\xspace}
\newcommand{\repoaudit}{RepoAudit\xspace}
\newcommand{\zerodays}{15\xspace}

\definecolor{DarkGreen}{rgb}{0,0.5,0}

\newcommand{\listingsttfamily}{\fontfamily{RobotoMono-TLF}\footnotesize}

\colorlet{punct}{red!60!black}
\definecolor{delim}{RGB}{20,105,176}
\colorlet{numb}{magenta!60!black}
\lstdefinelanguage{json}{
    basicstyle=\listingsttfamily,
    numbers=left,
    numberstyle=\scriptsize,
    stepnumber=1,
    numbersep=8pt,
    showstringspaces=false,
    breaklines=true,
    frame=lines,
    literate=
     *{0}{{{\color{numb}0}}}{1}
      {1}{{{\color{numb}1}}}{1}
      {2}{{{\color{numb}2}}}{1}
      {3}{{{\color{numb}3}}}{1}
      {4}{{{\color{numb}4}}}{1}
      {5}{{{\color{numb}5}}}{1}
      {6}{{{\color{numb}6}}}{1}
      {7}{{{\color{numb}7}}}{1}
      {8}{{{\color{numb}8}}}{1}
      {9}{{{\color{numb}9}}}{1}
      {:}{{{\color{punct}{:}}}}{1}
      {,}{{{\color{punct}{,}}}}{1}
      {\{}{{{\color{delim}{\{}}}}{1}
      {\}}{{{\color{delim}{\}}}}}{1}
      {[}{{{\color{delim}{[}}}}{1}
      {]}{{{\color{delim}{]}}}}{1},
}

\definecolor{myblue}{HTML}{4C85E5}
\definecolor{myred}{HTML}{D94357}
\definecolor{mygreen}{HTML}{4CAF50}
\lstdefinestyle{mystyle}{
  backgroundcolor=\color{white},   
  commentstyle=\color{mygreen},
  keywordstyle=\color{myblue},
  numberstyle=\tiny\color{gray},
  stringstyle=\color{myred},
  basicstyle=\listingsttfamily,
  breakatwhitespace=false,         
  breaklines=true,                 
  captionpos=b,                    
  keepspaces=true,                 
  numbers=left,                    
  numbersep=5pt,                  
  showspaces=false,                
  showstringspaces=false,
  showtabs=false,                  
  tabsize=2
}
\icmltitlerunning{VulnLLM}
\begin{document}
\twocolumn[
\icmltitle{VulnLLM-R: Specialized Reasoning LLM with Agent Scaffold for Vulnerability Detection}



\icmlsetsymbol{equal}{*}

\begin{icmlauthorlist}
\icmlauthor{Yuzhou Nie}{ucsb}
\icmlauthor{Hongwei Li}{ucsb}
\icmlauthor{Chengquan Guo}{uchicago}
\icmlauthor{Ruizhe Jiang}{ucsb}
\icmlauthor{Zhun Wang}{ucb}
\icmlauthor{Bo Li}{uiuc}
\icmlauthor{Dawn Song}{ucb}
\icmlauthor{Wenbo Guo}{ucsb}
\end{icmlauthorlist}

\icmlaffiliation{ucsb}{Department of Computer Science, University of California, Santa Barbara, CA, USA}
\icmlaffiliation{uchicago}{Department of Computer Science, University of Chicago, Chicago, IL, USA}
\icmlaffiliation{ucb}{Department of Electrical Engineering and Computer Sciences, University of California, Berkeley, CA, USA}
\icmlaffiliation{uiuc}{Department of Computer Science, University of Illinois Urbana-Champaign, Champaign, IL, USA}

\icmlcorrespondingauthor{Dawn Song}{dawnsong@berkeley.edu}
\icmlcorrespondingauthor{Wenbo Guo}{henrygwb@ucsb.edu}

\icmlkeywords{Machine Learning, ICML}

\vskip 0.3in
]
\printAffiliationsAndNotice{\icmlEqualContribution}
\begin{abstract}

We propose \sys, the~\emph{first specialized reasoning LLM} for vulnerability detection. 
Our key insight is that LLMs can reason about program states and analyze the potential vulnerabilities, rather than simple pattern matching.
This can improve the model's generalizability and prevent learning shortcuts. 
However, SOTA reasoning LLMs are typically ultra-large, closed-source, or have limited performance in vulnerability detection.    
To address this, we propose a novel training recipe with specialized data selection, reasoning data generation, reasoning data filtering and correction, and testing-phase optimization. 
Using our proposed methodology, we train a reasoning model with seven billion parameters.
Through extensive experiments on SOTA datasets across Python, C/C++, and Java, we show that \sys has superior effectiveness and efficiency than SOTA static analysis tools and both open-source and commercial large reasoning models.
We further conduct a detailed ablation study to validate the key designs in our training recipe. 
Finally, we construct an agent scaffold around our model and show that it outperforms CodeQL and AFL++ in real-world projects.
Our agent further discovers a set of zero-day vulnerabilities in actively maintained repositories.
This work represents a pioneering effort to enable real-world, project-level vulnerability detection using AI agents powered by specialized reasoning models.
The code is available at~\href{https://github.com/ucsb-mlsec/VulnLLM-R}{github}.
\end{abstract}

\section{Introduction}
\label{sec:intro}

The security community has a long history of using machine learning techniques for vulnerability detection, ranging from traditional ML models~\cite{risse2024uncoveringlimitsmachinelearning} to deep neural networks~\cite{zhou2019devign,Zou2019vul} and transformer-based models~\cite{du2024vulragenhancing}.
However, these approaches cannot be used in real-world applications due to multiple limitations.
First, the nature of most ML models is pattern learning and matching, constraining their generalizability to unseen programs and vulnerability patterns.
Second, existing ML models are small and have limited model capacity, which restricts them to analyzing only small and simple functions. 
As a result, traditional program analysis-based tools are still the SOTA techniques used in real-world vulnerability detection (e.g., CodeQL~\cite{codeql} and AFL++~\cite{AFLplusplusWoot20}).
\begin{figure}
    \centering
    \includegraphics[width=1\linewidth]{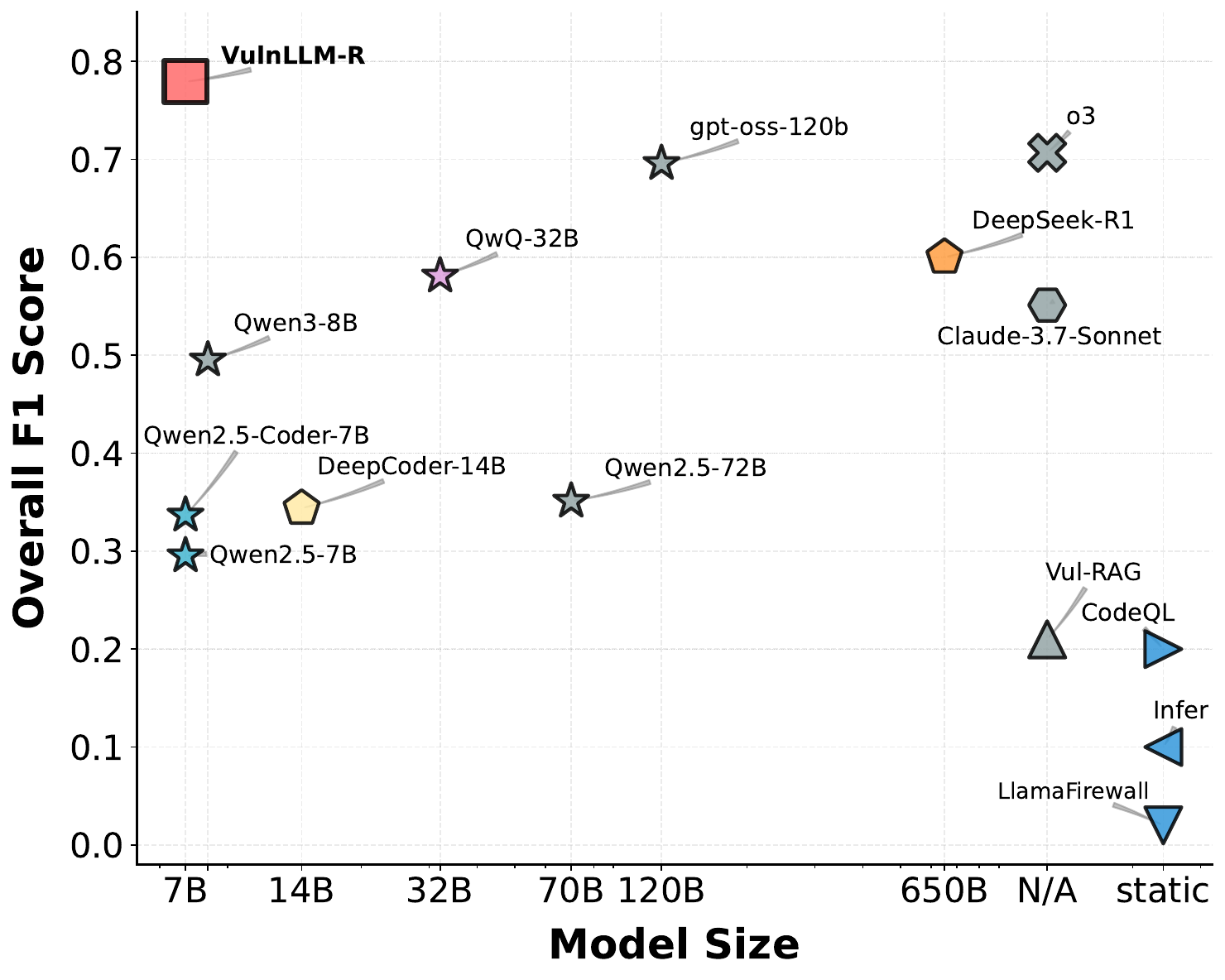}
    \caption{Performance vs. model size on vulnerability detection. \sys achieves SOTA performance with superior parameter efficiency.}
    \label{fig:model_f1_scatter}
\end{figure}

In this paper, we propose \sys, the~\emph{first reasoning LLM} specially designed for vulnerability detection. 
Different from traditional models that directly provide the required outputs, a reasoning model outputs its thinking process before giving the final outputs, where the model can predict the possible states of the input program and analyze whether the program contains a vulnerability.
Our motivations for training a specialized reasoning model rather than using SOTA reasoning models are threefold. 
First, existing reasoning models are ultra-large models serving general purposes, which have irrelevant capabilities (e.g., image and video processing), making them inefficient for a code-focused task -- vulnerability detection.
Second, detecting vulnerabilities requires understanding unique security vulnerabilities and principles, which general-purpose models may lack.
Finally, having in-house models avoids data privacy concerns, as well as supporting flexible integration and easier customizations.
While recent works have explored LLMs for vulnerability detection, they fall short in training specialized reasoning models.
For instance, LLM4Vul~\cite{sun2025llm4vuln} provides an evaluation framework for reasoning models but does not offer a training methodology.
LLMxCPG~\cite{lekssays2025llmxcpg} proposes a context-aware detection framework using code property graphs, yet trains only classification models rather than reasoning models.
Training reasoning models presents unique challenges compared to classification models: reasoning models must learn to articulate their analytical process, requiring carefully curated reasoning chains and specialized training strategies.
Without a systematic training recipe, simply scaling up model size or training data fails to produce effective reasoning capabilities for vulnerability detection.

Technically speaking, we distill large reasoning models by fine-tuning a small base model with supervised fine-tuning (SFT)~\cite{sky_t1_2025,muennighoff2025s1}.
Our~\emph{key technical contributions} lie in our novel training recipe, which includes data selection, reasoning data generation, reasoning data filtering and correction, and testing-phase optimization.
First, we select training data based on two criteria: \textit{CWE coverage and scale diversity}. Our selected samples include individual functions and vulnerabilities that involve multiple functions. 
Second, we use two SOTA open-source reasoning models as our teacher models: \dsr and \qwq.
This design can enable \sys to learn more diverse reasoning logic and structures, preventing model degeneration.
More importantly, we propose to filter out reasoning data (generated by teacher models) with wrong answers.
This is especially effective for vulnerability detection tasks, where the base small model does not have enough knowledge about security vulnerabilities.
Filtering out wrong answers helps our model learn the correct knowledge, not only the reasoning structure.
As the filtering process reduces the available training data, we further propose a~\emph{constitution-based correction}, where we write guidance for analyzing commonly wrong CWEs as additional instruction for teacher models.
Finally, we propose~\emph{summary-based training} to reduce our model's reasoning lengths and ~\emph{policy-based generation} to improve inference effectiveness.

We conduct a comprehensive evaluation on SOTA datasets across Python, C/C++, and Java, covering more than 50 CWEs. 
As shown in~\Cref{fig:model_f1_scatter}, our model can outperform many SOTA open-source and commercial reasoning LLMs with only a small model size (7B).
Besides, we also demonstrate the advantages of \sys's effectiveness in inference time.
We further show that \sys trained with a set of CWEs from Python and C/C++ can generalize to unseen CWEs and Java during inference, which cannot be observed in non-reasoning models.  
The poor performance of our base (non-reasoning) model validates that such generalizability comes from our customized training rather than the model's existing knowledge. 
Then, we conduct a comprehensive ablation study to validate the key designs in our training recipe. 
We also show that, different from general coding and math tasks, testing-phase scaling (long reasoning and parallel samplings) is not needed for our vulnerability detection tasks.  

Going beyond model learning, we further construct an agent that integrates \sys with CodeQL-based context retrieval. 
Given a large project, our agent first retrieves the necessary context (e.g., call paths and related functions) for each target function, and then feeds the function and its context to \sys to judge its vulnerability.
We continue to fine-tune our inference model under agent scaffolding to improve its ability to work with tools where agents provide valuable training data that is otherwise unavailable.   
We show that our agent outperforms \codeql~\cite{codeql} and \aflpp~\cite{AFLplusplusWoot20} and two LLM facilitated tools (\repoaudit~\cite{guo2025repoaudit}, \gtwofuzz~\cite{zhang2025low}) in 5 real-world projects.
Furthermore, we deploy our agent on the latest versions of 5 popular repositories and successfully discover \zerodays zero-day vulnerabilities, demonstrating the practical value of our approach in identifying previously unknown security issues in actively maintained software.

This paper makes the following contributions.

\begin{itemize}
    \item We propose \sys, the~\textbf{first specialized and small reasoning model} for vulnerability detection with a customized training recipe.

    \item We compare \sys with static analysis tools and a number of SOTA reasoning models and demonstrate its advantages in effectiveness and efficiency in~\Cref{fig:model_f1_scatter}. This is the first work that~\textbf{beat SOTA commercial models with specialized small models}, demonstrating the necessity and promise of developing specific reasoning models for security applications. 

    \item We construct the~\textbf{first reasoning model-based agent} and show its superiority over SOTA vulnerability detection tools, discovering \zerodays zero-day vulnerabilities in the latest repositories. It points to a promising new direction for vulnerability detection mechanisms where we construct agents with security-specific tools and train specialized models using the agentic data.  

\end{itemize}
\section{Background}
\label{sec:bg}

\subsection{Large Language Models}
\label{subsec:bg_llm}

LLMs are transformer-based neural networks~\cite{vaswani2017attention} with billions of parameters.
Given a sequence of input tokens $\mathbf{X} = [\mathbf{x}_1, ..., \mathbf{x}_L]$, LLM first predicts $\mathbf{x}_{L+1}$ based on the final layer representation of the last input token $\mathbf{o}_L$.
Then, it auto-regressively produces the next token $\mathbf{x}_{s+1}$ based on the current $\mathbf{o}_s$ until the end-of-sequence token is generated or it reaches an output limit. 
All of the generated tokens will be combined as the response for the current input. 
Benefiting from their ultra-high capacity and large training data, LLMs exhibit strong capabilities in understanding, reasoning, and content generation.
These capabilities enable LLMs to tackle complex tasks such as coding~\cite{chen2021evaluating}, solving math challenges~\cite{wang2023generative}, and conducting scientific discoveries~\cite{de2019synthetic}. 
Popular LLMs include closed-source models like OpenAI’s models~\cite{openai_gpt}, Google’s Gemini~\cite{google_gemini}, and Anthropic's Claude models~\cite{anthropic_claude}, as well as open-source models like Meta’s Llama models~\cite{touvron2023llama1,llama3}, DeepSeek models~\cite{guo2025deepseek}, and Alibaba Qwen models~\cite{yang2024qwen2,bai2023qwen,hui2024qwen2}.

\noindent\textbf{Training.}
LLMs training typically has two stages: pre-training and fine-tuning.
Pre-training conducts self-supervised-based next-token prediction with the following objective function $  \sum_{i=2}^{L} log(P(\mathbf{x}_i \mid \mathbf{x}_1, \mathbf{x}_2, \ldots, \mathbf{x}_{i-1}; \Theta))$.
This process allows the model to learn general knowledge~\cite{radford2018improving,radford2019language}.
The fine-tuning process has two possible training methods: supervised fine-tuning (SFT) and reinforcement learning (RL).
The goal for this stage is to calibrate the model for specific tasks or safety alignment~\cite{ouyang2022training}.
SFT applies the next token prediction to a labeled dataset.
RL has offline and online methods, where offline methods (e.g., Direct Preference Optimization (DPO)~\cite{rafailov2024direct}) require data sampled from an expert model/policy, and online methods (e.g., Proximal Policy Optimization (PPO)~\cite{schulman2017proximal}) need a reward function to assign a reward to the model output.
The typical RL learning objective is to maximize the expected total reward while maximizing the policy entropy. 

\noindent\textbf{Inference.}
Users interact with a trained LLM by providing \textit{input prompts}.
For example, it could be a question {``How to create a file in a Linux system?"}.
Recent research explores \textit{in-context learning}~\cite{brown2020language}, which uses \textit{few-shot examples} to help the LLM understand the input prompt. 
Here, the few-shot example could be {``How to remove a file in a Linux System? Answer: rm -rf file\_name"}.
To gain global control over the model’s outputs, developers often provide a \textit{system prompt}, which provides global guidelines and assumptions for the model's responses. 
For instance, a system prompt could be {``You are a computer science expert who can answer users' questions correctly."}
This system prompt is typically placed before the user input prompt, e.g., with the system prompt, the above prompt becomes {``You are a computer science expert who can answer users' questions correctly. How to remove a file in a Linux System? Answer: rm -rf file\_name. How to create a file in a Linux system?"}.

\subsection{Reasoning LLMs}
\label{subsec:bg_reasoning}

\noindent\textbf{Testing-phase reasoning.}
Going beyond few-shot learning, recent works develop more advanced prompting strategies that can stimulate the model to output its \textit{reasoning process} while responding to an input query. 
For example, Chain-of-Thought (CoT)~\cite{wei2022chain} adds the instruction of ``Please think step by step" in the prompt, which will guide the model to output its reasoning process. 
Follow-up works propose more advanced prompting methods to instruct the model to do deeper and more complex reasoning (e.g., Tree-of-Thought~\cite{yao2023tree}, ReAct~\cite{yao2023react}).
With the reasoning process, LLMs can better analyze the input queries, conduct deeper thinking, and even justify their decisions rather than merely replaying learned patterns.
These studies show that the ability of LLMs on complex tasks such as code generation and mathematical solving is significantly improved thanks to the reasoning process.

\noindent\textbf{Train a reasoning model.}
Motivated by the superior performance of the additional reasoning process, researchers further explore training a reasoning model that teaches the model to do reasoning during the fine-tuning stage.
Such models typically have the reasoning instruction in the system prompt (e.g., ``Please think step by step'').
There are two lines of work based on the fine-tuning methods introduced above.  
First, we can train a model to distill another reasoning model using SFT.
That is, we craft a dataset where the response of each input has an explicit reasoning process.
The data are generated by another reasoning LLM~\cite{singh2023beyond,muennighoff2025s1,li2025llms,ye2025limo}.
Then, we fine-tune the model with this dataset, which forces the model to mimic the training data and conduct reasoning. 
Second, recent work shows that LLMs can automatically learn to conduct reasoning when trained with online RL methods using a final outcome reward (i.e., a reward assigned at the last token indicating whether the answer is correct)~\cite{xiong2025self,arora2025training,guo2025deepseek}. 
Follow-up works either explore improving the RL algorithm under this outcome reward~\cite{yu2025dapo,yuan2025vapo}, learning a process reward model (PRM) that assigns every token or reasoning step a reward~\cite{uesato2022solving,lightman2023let,zhao2025genprm,zhang2024entropy,wang2023math,zhang2024rest}, or explore offline RL methods for training reasoning models~\cite{rafailov2024r,zeng2024token,ji2024enhancing,wang2024offline}.  
All these reasoning models show better performances in code-related tasks, such as code generation and code understanding.
Besides, some works further show that having a long reasoning chain during inference can further improve the model's coding capabilities (\textit{testing-phase scaling})~\cite{muennighoff2025s1,li2025llms}. 
The State-of-the-art (SOTA) reasoning LLMs include closed-source models, like \omini~\cite{o4-mini} and \claude~\cite{claude37}, and open-sourced models, like \qwq~\cite{QwQ32B} and \dsr~\cite{guo2025deepseek}. 

\begin{figure*}
    \centering
    \includegraphics[width=1\linewidth]{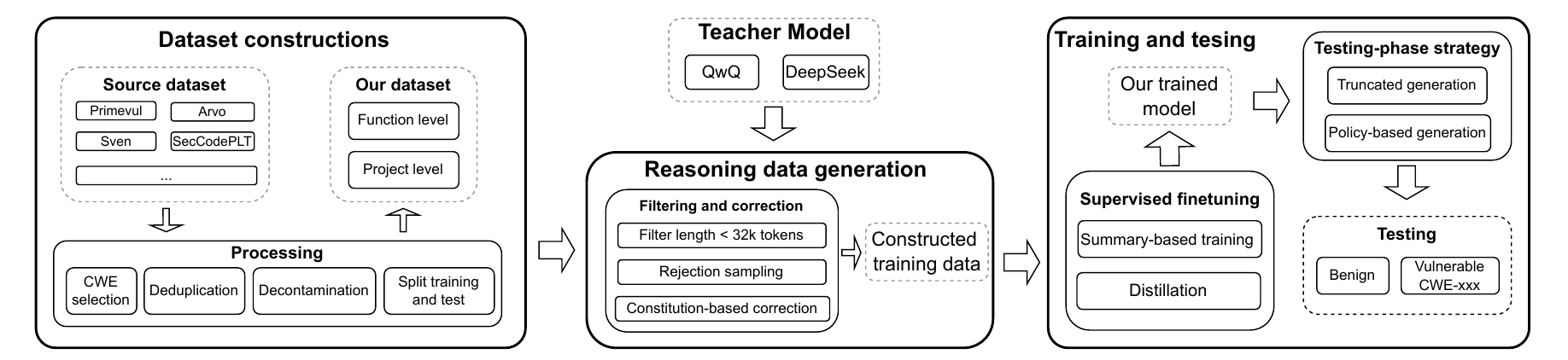}
    \caption{Our proposed training recipe, including training data construction, training methods, and testing strategies.}
    \label{fig:overview}
\end{figure*}

\section{Overview of \sys}
\label{sec:tech}

Motivated by the impressive performance of reasoning models in general coding tasks, we explore whether reasoning capabilities can be migrated to vulnerability detection, a task that requires understanding the target program and critical security principles.
Following this idea, we design \sys, the~\emph{first} specialized reasoning model for vulnerability detection.
We fine-tune an open source LLM into a reasoning model with our~\emph{proposed training recipe}.
We carefully optimize each step in the recipe, including data selection, reasoning data generation and filtering, and inference-phase optimization, to minimize the model size while maximizing the model performance.
In what follows, we specify our problem definition and high-level insights for our training recipe.
\Cref{sec:tech_details} further provides more details.

\subsection{Problem Scope}
\label{sec:tech_scope}

\noindent\textbf{Programming languages and vulnerabilities.}
We consider four widely used programming languages: Python, C, C++, and Java, which account for most of the common security vulnerabilities.
They also cover three major programming paradigms: process-oriented (C), object-oriented (C++ and Java), and scripting (Python).
As shown in~\Cref{tab:dataset-composition}, for each set of languages, Python, C/C++, and Java, we select a group of common CWEs covered by existing benchmarks.
We exclude CWEs that are of low severity, rarely observed, or have very few data points in existing benchmarks.
We use data from the selected CWEs to train our model and test it on a different set of data belonging to the same CWEs.
Furthermore, we also hold out an out-of-distribution (OOD) testing set, with unseen CWEs during training.

\noindent\textbf{Program complexity: function-level and project-level.}
Most of the existing ML-based approaches and vulnerability detection benchmarks focus on the function level.
That is, given as input the individual functions, the model decides whether each function is vulnerable or not.
In this paper, we start with the same function level to better leverage the existing benchmarks for training and testing.
However, we also go beyond the individual function level and explore the possibility of conducting vulnerability detection for an entire project.
Although this will introduce additional challenges in model training and deployment, pushing ML-based vulnerability detection to the project level marks a notable milestone towards their applications in real-world security problems.

Under this setup, our goal is to train a~\emph{small reasoning model that can correctly identify the vulnerabilities in input functions or projects, even when the vulnerability belongs to OOD CWEs.}
\emph{We further design the model to not only detect the vulnerability but label it with the correct CWE, which is valuable context for triage and patching.}

\subsection{Technical Overview}
\label{sec:tech_overview}

\noindent\textbf{Rationale for specialized reasoning models.}
We first explain our insights for developing a vulnerability detection reasoning model from two aspects: 1) why we need reasoning models; 2) why we need to train a specialized model rather than using existing commercial models.

\noindent\underline{Reasoning models.}
As introduced in~\Cref{subsec:bg_reasoning}, a reasoning model can automatically output its internal reasoning process and then give the final answer based on the reasoning.
The reasoning process enables the model to analyze the input and thus better solve complex tasks requiring deep thinking.
Reasoning capabilities are necessary for vulnerability detection for the following two reasons.
First, an ideal vulnerability detection process inherently requires deep reasoning and thinking, even for human experts, where we need to understand the input program, reason about its program states, and decide whether the code belongs to a certain vulnerability.
These attributes make the reasoning model a good candidate for solving vulnerability detection tasks.
Second, without an explicit reasoning process, the model may degrade to pattern learning.
It is easy for such a model to~\emph{learn shortcuts in the input code and thus have limited generalizability}~\cite{risse2024scorewrongexambenchmarking}.
Third, reasoning capabilities provide better out-of-distribution (OOD) generalization across different programming languages and unseen CWE types.
By reasoning about program semantics and security principles rather than memorizing patterns, the model can better transfer vulnerability detection knowledge across languages, as many vulnerabilities share similar underlying security logic.

\noindent\underline{Specialized reasoning models.}
We argue the necessity of developing specialized reasoning models for vulnerability detection from the following three aspects.
First, most SOTA LLMs are general-purpose models encoding multiple capabilities and thus have a very large size.
However, many capabilities are useless for our task, such as solving math problems or processing images.
Second, vulnerability detection requires unique knowledge about the security principles and practices, which may not be comprehensively acquired by general-purpose models.
As shown in~\Cref{sec:eval_main}, SOTA models, though achieving very high performance in coding tasks~\cite{openai_gpt,claude37}, still fail to identify many vulnerabilities, which indicates a capability gap.
As such, we believe it is necessary to train~\emph{small but specialized} models that do not need to solve math problems or process images; instead, reason about the input code and precisely identify vulnerabilities.
Such a model can also be more cost-efficient during deployment and can be easier to update and adapt to new vulnerabilities and use cases.
Finally, in-house models address the concerns about data privacy when using a third-party model.
Having full control of the models also makes it easier for developers to construct agent scaffolds with more flexible interactions between models and other components.

\noindent\textbf{Our training recipe in a nutshell.}
We propose to train our model by distilling large reasoning models with supervised fine-tuning (SFT), as it enables training effective small reasoning models with limited computational resources~\cite{sky_t1_2025,muennighoff2025s1}.
In contrast, training reasoning models with reinforcement learning requires substantially more computational power.
Moreover, if training is guided only by an outcome reward, such as whether the final judge of vulnerability is correct, and lacks well-designed intermediate signals, it becomes unstable and highly susceptible to reward hacking~\cite{amodei2016concrete}.

Due to the task differences, we cannot directly apply the existing distillation recipes designed for general coding and math problems; instead, we need to design a customized recipe.
~\Cref{fig:overview} shows the key novelty and uniqueness in our training recipe.

First, we select training data based on two criteria: \textit{CWE coverage and scale diversity}.
We select both vulnerable and benign data that cover multiple CWEs written in different languages.
For benign data, we include the patched code of the vulnerable ones to let the model distinguish between the vulnerable and secure code.
We also include some normal code snippets to enable the model to learn about common properties.
For scale diversity, we include data with \emph{different scales}, measured by their line of code (LoC).
We extract both short functions from the \juliet dataset~\cite{black2018juliet} and some long functions from the \primevul dataset~\cite{ding2024primevul}.
We further include samples from the \arvo dataset~\cite{mei2024arvo}, where each vulnerability may involve multiple functions or even multiple files (e.g., ARVO 15374 involves 6 different cpp files).
This scale diversity enables the model to digest and handle different situations during training.
Besides, as longer code typically means more difficult cases, as shown in recent works~\cite{muennighoff2025s1}, having difficult data in the training set can improve the model performance.
We further deduplicate and decontaminate our training data to remove duplicate samples and prevent test set leakage, ensuring the integrity of our evaluation.

Second, we propose to use two SOTA open-source reasoning models to generate reasoning data: \dsr and \qwq.
Our motivation for choosing these open-source models, rather than other alternatives, is primarily driven by the goal of demonstrating that a reasoning model fine-tuned with data from open-source models can surpass commercial models.
As demonstrated in~\Cref{sec:eval_ablation}, these two models have different reasoning logic and structure and thus have different biases towards false positives and false negatives.
As such, mixing their reasoning data in the training set can balance our models in controlling both errors.
Besides, having more diverse data generated from two SOTA models can prevent our model from degeneration during the SFT training (i.e., the model iteratively outputs the same content).

Third, we propose a novel reasoning data filtering and correction method.
We follow existing works and filter out data with a very long reasoning chain (total tokens $\geq$ 32,000), as they typically include redundant reasoning steps and will make the training inefficient.
We further filter out the reasoning data with incorrect final answers (\emph{rejection sampling}).
This is~\textit{different from existing works} (e.g., S1~\cite{muennighoff2025s1}), which show no need for rejection sampling in the general coding and math problems, as such data is still useful for models to learn the reasoning structure.
We believe the difference is that a small model needs to learn the correct knowledge during fine-tuning, not only the reasoning structure.
In contrast, the model has likely seen enough examples in pre-training for general tasks and can tolerate occasional wrong answers.
However, applying rejection sampling will significantly reduce the available number of training samples.
To address this, we propose~\emph{constitution-based correction}, which re-queries the teacher models using failure cases.
We analyze the typical failures of training CWEs and manually construct constitutions as additional instructions that guide the model to better analyze the corresponding CWEs.
For example, when analyzing path traversal vulnerabilities (CWE-22), teacher models may fail to identify subtle bypasses like normalized paths or symbolic links.
Our constitution addresses this by instructing the model to ``verify path canonicalization, check for directory traversal sequences, and confirm proper sandboxing within allowed directories''.


We then propose~\emph{summary-based training}, where we query the teacher models to summarize their own reasoning data and further calibrate our model through SFT on these samples.
Our empirical results show that this additional calibration training can guide the model to give concise answers without sacrificing performance.
As shown in~\Cref{app:comparedpo}, this approach is better than DPO-based offline RL in our problem.

During the testing phase, we propose two strategies:~\emph{truncated generation} to improve inference efficiency and \emph{policy-based generation} to reduce task complexity.
Truncated generation forces the reasoning process to stop and output the final answer once the reasoning chain reaches a certain length.
Given an input sample, policy-based generation first queries the model 4 times and uses its outputs as a set of potential CWE candidates, denoted as policy.
Then, we add the policy to the input and only query the model to decide whether the input belongs to one of the CWEs in the policy.

\subsection{Agent Scaffold of \sys}
\label{sec:tech_agent}

Our training recipe described above yields a small model that, when supplied with sufficient context, excels at reasoning and detecting vulnerabilities.
Scaling the model to full‑project analysis, however, demands an automated way to gather the required context—something rule‑based static tools can’t reliably provide.
To solve this, we wrap our reasoning model in a lightweight agent scaffold with a context‑retrieval component.
At a high level, we apply our model to every function in the project, allowing it to retrieve the context it needs autonomously.
For each function, we first extract all functions along three randomly sampled paths from the project's entry point to this function in the call graph, providing these functions as the initial context to the model.
We select three paths to balance the trade-off between providing sufficient initial context and staying within \sys's input length constraints.
Additionally, we equip the model with a tool capable of retrieving function implementations by name, allowing it to autonomously obtain further context on demand, while restricting the number of interaction rounds.
We adopt this design specifically because, at the project level, we typically need to perform vulnerability detection over thousands of functions; thus, controlling inference speed through limited rounds is essential.
As shown in~\Cref{app:context}, such context provides critical program artifacts necessary for model decisions, without which, the model may introduce additional false positives and false negatives.

While the training recipe described above produces a model capable of vulnerability detection reasoning, it is not specifically trained to use context retrieval tools, which are essential for project-level analysis.
To address this limitation, we continue training our model using agentic traces generated by the \othree\cite{o4-mini} model with the agent scaffold described above.
We choose \othree rather than our two open-source teacher models because it is better at using tools.
We select several commonly used C/C++ and Java repositories from GitHub to construct our training set, including Assimp~\cite{assimp2025}, SQLite3~\cite{sqlite2025}, CUPS~\cite{cups2025}, and libxml2~\cite{libxml2} for C/C++, as well as zt-zip~\cite{zt-zip} and Apache Commons Compress~\cite{commons-compress} for Java.
For each repository, we manually integrate multiple vulnerabilities corresponding to that repository from the \arvo dataset, creating realistic project-level vulnerability detection scenarios.
The details of each repository and its associated vulnerabilities are presented in~\Cref{app:agent_data_targets}.
As shown in~\Cref{sec:application}, training the model using agentic data is a novel and efficient way of learning tool calling capabilities, has recently begun to be used in other fields as well~\cite{goutora,golubev2025training,ma2025tool,feng2025retool}.
We use \othree to collect data, employing rejection sampling to discard samples with incorrect vulnerability judgments.


\section{Technical Details}
\label{sec:tech_details}

\subsection{Dataset Collection}
\label{sec:data_collection}

\begin{table*}[t]
  \centering
  \caption{Composition of our dataset, grouped by training and testing sets.
  }
  \label{tab:dataset-composition}
  \resizebox{0.98\textwidth}{!}{
  \begin{tabular}{@{}rrlcccccc@{}}
    \toprule
    \textbf{} & \textbf{Source} & \textbf{Scale} & \textbf{Language} & \textbf{CWE} & \textbf{\# Benign} & \textbf{\# Vuln.} & \textbf{Avg. Length} \\
    \midrule
    \multicolumn{8}{l}{\textit{Training Set}} \\
    \midrule
    & \seccodeplt\cite{yang2024seccodeplt},  \sven\cite{he2023large}     & Function Level         & Python & 18 & 1281 & 1281 & 741 \\
    & \sven\cite{he2023large}, \juliet\cite{black2018juliet}, \primevul-filtered\cite{ding2024vulnerability}     & Function Level         & C/C++  & 31 & 4736 & 4497 & 4484 \\
    & \secllm\cite{ullah2024llms}, \arvo\cite{mei2024arvo}      & Project Level  & C/C++  & 3  & 568  & 604  & 13415 \\
    \midrule
    \multicolumn{8}{l}{\textit{Testing Set}} \\
    \midrule
    & \seccodeplt\cite{yang2024seccodeplt},  \sven\cite{he2023large}     & Function Level        & Python & 24 (6 ood)  & 71  & 71  & 795 \\
    & \sven\cite{he2023large}, \juliet\cite{black2018juliet}, \primevul-filtered\cite{ding2024vulnerability}         & Function Level        & C/C++  & 45 (14 ood) & 775 & 507 & 3548 \\
    & \juliet\cite{black2018juliet}      & Function Level  & Java  & 25   &  2727  & 3082 & 2663 \\
    & \secllm\cite{ullah2024llms}, \arvo\cite{mei2024arvo}      & Project Level  & C/C++  & 3   & 198   & 320 & 19315 \\
    \bottomrule
  \end{tabular}
  }
\end{table*}

\noindent\textbf{Dataset construction.}
As discussed in~\Cref{sec:tech}, our data selection strategy emphasizes both CWE coverage and scale diversity. 
Guided by these criteria, we choose six widely used datasets, including \juliet~\cite{black2018juliet}, \arvo~\cite{mei2024arvo}, 
\secllm~\cite{ullah2024llms}, \primevul~\cite{ding2024vulnerability}, \sven~\cite{arendt2014sven}, and \seccodeplt~\cite{yang2024seccodeplt}. 
These datasets are selected not only because they are commonly used benchmarks in vulnerability research, but also because they collectively cover a broad range of CWE types and programming languages, and span different data granularities (i.e., some contain only function-level information, while others provide project-level contexts).

For \secllm~\cite{ullah2024llms} and \sven~\cite{arendt2014sven}, we include all samples in our selected CWE list.
For \juliet, we select the long samples, which tend to provide more complexity.
For \primevul and \seccodeplt, some samples lack sufficient context to reliably determine their labels, and some samples even have wrong labels.
We design an LLM-assisted human validation process to filter out wrong or ambiguous samples. 
For \arvo, we only retain samples whose sanitizer output clearly matches CWEs within our defined scope.
The original benign samples in the selected datasets are mostly patched versions of the vulnerable code.
To let our model learn the behaviors of normal benign programs, we further include normal functions from the \primevul-nopair dataset (the standalone benign functions without vulnerable counterparts from PrimeVul). 
Details of the selected CWEs and our benchmark-specific filtering operations are provided in~\Cref{app:sample_selection}.

To prevent data leakage caused by overlapping vulnerabilities across different datasets, we perform de-duplication using a 20-token n-gram matching method.
Specifically, we remove samples that contain overlapping sequences of at least 20 consecutive tokens between the training and test sets, and we apply the same criterion to eliminate such overlaps within the training and test sets individually.

\Cref{tab:dataset-composition} shows the composition of our final dataset, where the differences in average sample length indicate a diversity in difficulty.
\Cref{tab:dataset-composition} also shows the CWE coverage, where we prioritize CWEs overlapping with the 2024 CWE Top 25 Most Dangerous Software Weaknesses~\cite{top25cwe}.
Besides the above per-benchmark filtering, we also exclude CWEs where manual analysis reveals most samples are mislabeled. 
\noindent\textbf{Sample construction.}
Each sample in our dataset consists of two elements: a code snippet and a ground-truth label. 

\noindent\underline{The code snippet} is the piece of code for which the model must detect the presence or absence of a vulnerability.
For most datasets~\cite{yang2024seccodeplt,ding2024vulnerability,ullah2024llms}, each sample already contains a code segment labeled as either vulnerable or benign, along with the corresponding CWE number; we directly use the original code. 
For each sample in the \arvo dataset, vulnerable and patch commits from a GitHub repository are provided, together with a proof-of-concept (PoC) that triggers the vulnerability and causes a crash. 
We execute the PoC and extract all functions from the stack trace, as well as all functions modified by the patching commit. 
We then concatenate these functions to form the code snippet. 
In the prompt, we explicitly mark the functions modified by the patch commit as target functions that should be examined for potential vulnerabilities. 
Functions appearing in the stack trace are marked as context functions, providing relevant execution context but not being the primary focus of vulnerability reasoning.
For \juliet, each sample is a compilable program that typically contains both vulnerable and benign implementations. 
By setting environment variables during compilation, we can only include the vulnerable or the benign implementations and generate either a benign or a vulnerable version of the program.
We parse the source code of each program and extract the relevant code sections for each version, remove comments, and sanitize identifiers, such as namespace names, function names, and print statements, to avoid leaking ground-truth label information.
Each version is then stored as the code snippet of an individual sample in our dataset.
Further details of data processing are described in~\Cref{app:preprocess_data}.

\noindent\underline{The ground-truth label} indicates whether the code snippet is vulnerable; if so, the label also includes the corresponding CWE number. 
For the \arvo dataset, we infer the CWE number by pattern-matching the sanitizer output from the vulnerable program crash (details in~\Cref{app:preprocess_data}); for \secllm, we manually annotate each sample's CWE number.
For other datasets, each sample already comes with a CWE number, which we adopt as the ground-truth CWE number. 

\subsection{Training and Inference}
\label{sec:reasoning_generation}

\noindent\textbf{Base and Teacher models.}
As mentioned in~\Cref{sec:tech}, we generate reasoning-chain data from two SOTA open-source reasoning models, \dsr and \qwq. 
This choice enhances the diversity of reasoning structures and ensures broad coverage across different CWE types. 
We select \qweninstruct as our base model.

\noindent\textbf{Reasoning data filtering.}
First, we filter out samples if the combined length of input and output tokens exceeds $32,000$.
This threshold is set because the default maximum context length of our base model is $32,000$ tokens, and reasoning chains longer than this limit will cause degradation in model performance.
Second, we find that only a small fraction of samples receive correct final answers from at least one model when we generate reasoning chains from \dsr and \qwq.
It leaves the majority of the dataset without usable reasoning chains. 
To alleviate this limitation, we employ the following strategy:
\ding{192} \textbf{Rejection sampling}: For each data point, we query both \dsr and \qwq 8 times and retain the reasoning data that (i) leads to the correct final answer and (ii) has the shortest reasoning chain.
\ding{193} \textbf{Constitution-based correction}: For data where all 8 attempts fail, 
we further analyze representative errors and create constitutions for the models that summarize the general rules for identifying each CWE.
We then re-query the model and apply our rejection sampling strategy again, where the additional information (constitutions) can improve the teacher model's accuracy.  
We create 73 constitutions for 19 CWEs. 
The human-written constitutions are detailed in~\Cref{app:prompt_template}.
Through the constitution-based correction, we can obtain around 30\% more training data with the correct answers. 

\noindent\textbf{Training data construction.}
Each training sample consists of two components: the input prompt and the reasoning trajectory. 
During the distillation process, we query teacher models (\dsr and \qwq) to generate reasoning trajectories for each code snippet.
As described above, we may include some additional information in the input prompt when querying teacher models.
When constructing the final training data for our model, we remove it from the input prompt, as the testing phase does not provide guidelines.
The training data thus consists of the code snippet as input and the teacher-generated reasoning trajectory as output.
We provide the prompt details in~\Cref{app:prompt_template}.

\noindent\textbf{Summary-based training.}
We find that teacher models may generate reasoning chains with redundant steps, such as unnecessary repeated double-checking or analysis of irrelevant details. 
To address this, we apply a \emph{summary-based training} strategy. 
Specifically, we use \dsr to summarize the original reasoning chains, selecting only important and relevant steps while preserving the overall reasoning structure. 
We first fine-tune the base model on the original, full reasoning chains, then further fine-tune it on the newly summarized chains. 
This two-stage approach helps the model acquire comprehensive reasoning skills and also encourages it to perform reasoning in a more concise and efficient manner.

\noindent\textbf{Inference: Truncated generation.}
We employ truncated generation to balance efficiency and accuracy.
Motivated by S1~\cite{muennighoff2025s1}, we limit the reasoning length by appending a ``Final Answer'' token after a fixed amount of reasoning, forcing the model to conclude based on its partial analysis.

\noindent\textbf{Inference: Policy-based generation.}
First, we define a \emph{policy} as a set of CWEs that restricts the model's output to one of the identified CWE candidates.
We initially sample the model $4$ times with truncated reasoning to identify $2\sim 5$ distinct potential CWE candidates as our generated policy.
We then add this generated policy to the input prompt and prompt the model to make a final prediction.
This two-stage approach can help the model reason more accurately within the constrained scope, but at the cost of longer inference time.

\subsection{Failure Attempts and Analysis}
\label{sec:failure}

\noindent\textbf{Data Collection.}
First, we directly use the original \primevul training set, which leads to poorly performing models. 
This guides us to find the main issues mentioned above: lack of necessary context and incorrect labels.
Second, we attempt to use only patched functions as benign samples. 
This strategy makes it easy for our model to learn shortcuts, as the patched code differs by only a few lines from the vulnerable code (in most datasets, e.g., \primevul, just one additional assertion line).
Moreover, the patched code is sometimes not secure as the patches are not sound, which confuses the model into making incorrect decisions on actual benign code.
As such, we include normal programs, such as simple utility functions or well-tested core components, as additional benign programs in our training set.
Finally, we also explore training the model jointly on a general coding task.
Specifically, we use the reasoning data generated from Sky-T1-17k~\cite{sky_t1_2025}.
Sky-T1-17k contains math, coding, and science fields, and we find that using the entire dataset or only the coding task cannot help the model learn reasoning for vulnerability detection.
We suspect this is because the model is already pre-trained on similar general coding data, so the additional training does not add useful new information for this task.

\noindent\textbf{Reduce inference length (via offline RL).}
We try the following alternatives to our summary-based training. 
First, we prompt teacher models to give short reasoning (``think for up to $N$ tokens'' or ``do not spend excessive time double-checking your work''). 
The models do not follow the instructions and still give long reasoning. 
We then use budget forcing~\cite{muennighoff2025s1}, which indeed forces the teacher models to give short reasoning. 
However, it commonly truncates reasoning chains in ways that break the logical structure of the reasoning process. 
This leads to a significant decline in model performance, as our model cannot learn the complete reasoning structure.
However, we observe in~\Cref{sec:eval_testing} that truncating the reasoning during the testing phase can even improve performance.
We believe this is because the model already learns to perform reasoning, and testing-phase truncation can make it reason more efficiently.

We also apply the offline RL method DPO~\cite{rafailov2023directpreference} to further optimize our fine-tuned model towards providing shorter reasoning.
Specifically, we generate paired data by sampling reasoning chains of different lengths from the same input data, and then train the model to prefer shorter reasoning chains that still yield correct answers.
However, this approach does not yield significant improvements in shortening reasoning length or model performance.
We suspect that the limited diversity in reasoning chains generated by a single model may account for this failure, as the model does not encounter enough varied reasoning structures to learn effectively, and it is prone to generating collapse samples (e.g., repetitive or degenerate outputs).

\noindent\textbf{Online RL after distillation.}
Following existing works, we further train our model with online RL, such as GRPO~\cite{deepseek-math} and DAPO~\cite{yu2025dapoopensourcellmreinforcement}, with an outcome-based reward function.
Our reward design consists of four score levels: the highest score requires correctly identifying both vulnerability presence and CWE type; intermediate scores are given for correctly identifying vulnerability presence only; lower scores are given for completely incorrect predictions; and the lowest score is given for incorrect formatting. 

We use the samples for which the teacher models cannot provide correct answers for RL training, given that they represent more difficult cases.
We use the GRPO algorithm and find performance degradation after the RL training. 
Specifically, the false negative rate increases dramatically, meaning the model becomes more inclined to predict an input as benign. 
We hypothesize this is because of reward hacking. 
When the model fails to answer correctly in all $n$ queries in each round, the normalized reward becomes zero, providing no reward or penalty to the model. 
When the model predicts vulnerability presence, it must also correctly identify the CWE type to achieve the highest reward; otherwise, it will get penalized and receive a lower reward.
This is a much more difficult task than simply outputting benign and receiving zero reward. 
As such, the model learns this lazy behavior during the RL training by exploiting the design flaws in the reward function.
Given the non-trivial effort needed, we defer RL training with a proper reward function to future work.

\section{Evaluation}
\label{sec:eval}

\begin{figure*}[th]
    \centering
    \includegraphics[width=1\linewidth]{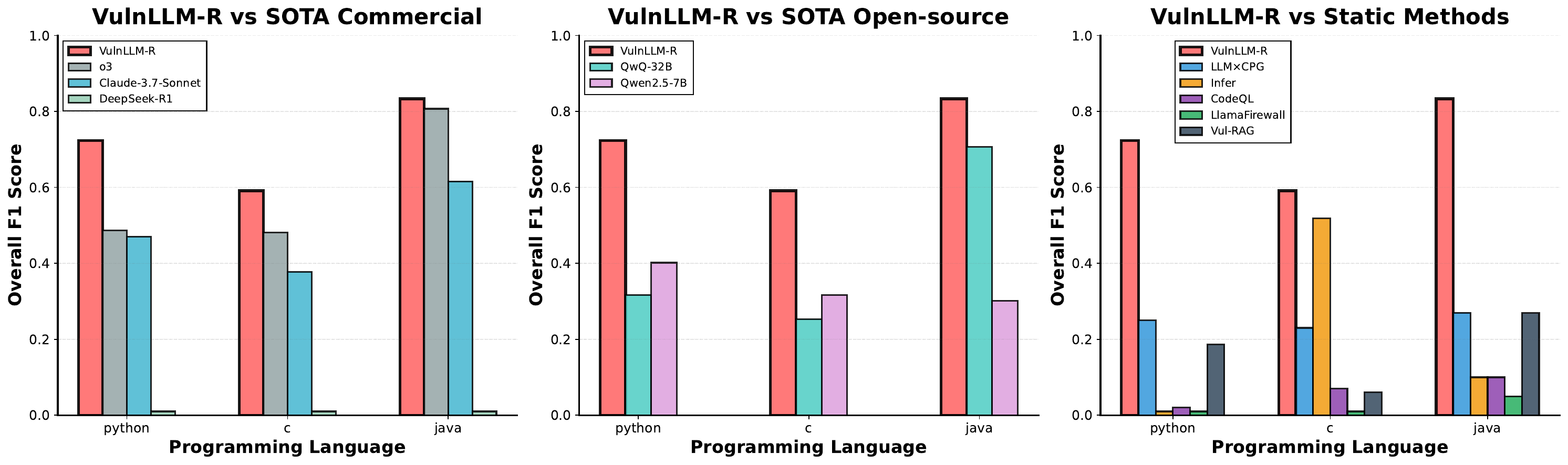}
    \caption{\sys vs. selected baselines on the testing set of individual languages (F1 Score).}
    \label{fig:main}
\end{figure*}

In this section, we evaluate \sys from the following aspects.
First, we show \sys's advantage over SOTA static analysis tools, commercial and open-source LLMs.
Second, we conduct a detailed ablation study to validate our key designs: reasoning models, distilling from two models, reasoning data filtering and correction, and summary-based training.
Third, we test \sys's performance on testing-phase scaling (output with different reasoning lengths and number of parallel queries).
Finally, we evaluate the sensitivity of \sys to three key hyperparameters: the number of queries to generate policy, the base model family, and model size in~\Cref{app:hyper-parameter}.
We use our dataset described in~\Cref{sec:data_collection} for this evaluation, which merges existing function-level benchmarks with the necessary filtering of noisy data.
Our dataset can be used as a clearer and more comprehensive benchmark for future works on vulnerability detection.


\subsection{\sys vs. Baseline Approaches}
\label{sec:eval_main}

\noindent\textbf{Setup and design.}
We select two SOTA static analysis tools: CodeQL~\cite{codeql} and Infer~\cite{infer}.
We also include LlamaFireWall~\cite{llamafirewall}, a recently released guardrail system with a rule-based vulnerability detector.
For LLM-based vulnerability detection methods, we evaluate LLMxCPG~\cite{lekssays2025llmxcpg}, and a RAG-based method, Vul-RAG~\cite{du2024vulragenhancing}, with \gptfourone as the model.
To ensure a fair comparison with LLMxCPG, we train it on our training data using their approach: Joern-based taint analysis to extract context.
Regarding purely LLMs, we compare \sys with its base model \base~\cite{yang2024qwen2}, two teacher models \dsr~\cite{guo2025deepseek} and \qwq~\cite{QwQ32B}, and two commercial models \othree~\cite{o4-mini} and \claude~\cite{claude37}. 
We query all the selected models with the same prompts.
We compare \sys with the baseline methods in identifying the correct CWEs (F1 score, false positive rate, and false negative rate) as well as their average testing-phase runtime on one sample.  

\noindent\textbf{Results.}
The results first show that static analysis tools perform worse than LLMs, demonstrating the effectiveness of LLMs over traditional rule-based approaches.
LLMs can learn broad knowledge and, more importantly, reason about inputs rather than simple pattern matching, which enables them to better analyze the input code, choose what knowledge to apply, and even criticize and refine their own judgment.
Notably, LLMxCPG shows significantly weaker performance than \sys despite our fair comparison setup. This stems from two fundamental differences: (1) Our method captures complete context while their taint-analysis-based extraction may miss critical information, and (2) LLMxCPG is a pure classification model that outputs a single word, whereas \sys is a reasoning model with superior understanding capabilities on complex problems.

\begin{figure}
  \centering
  \includegraphics[width=1\linewidth]{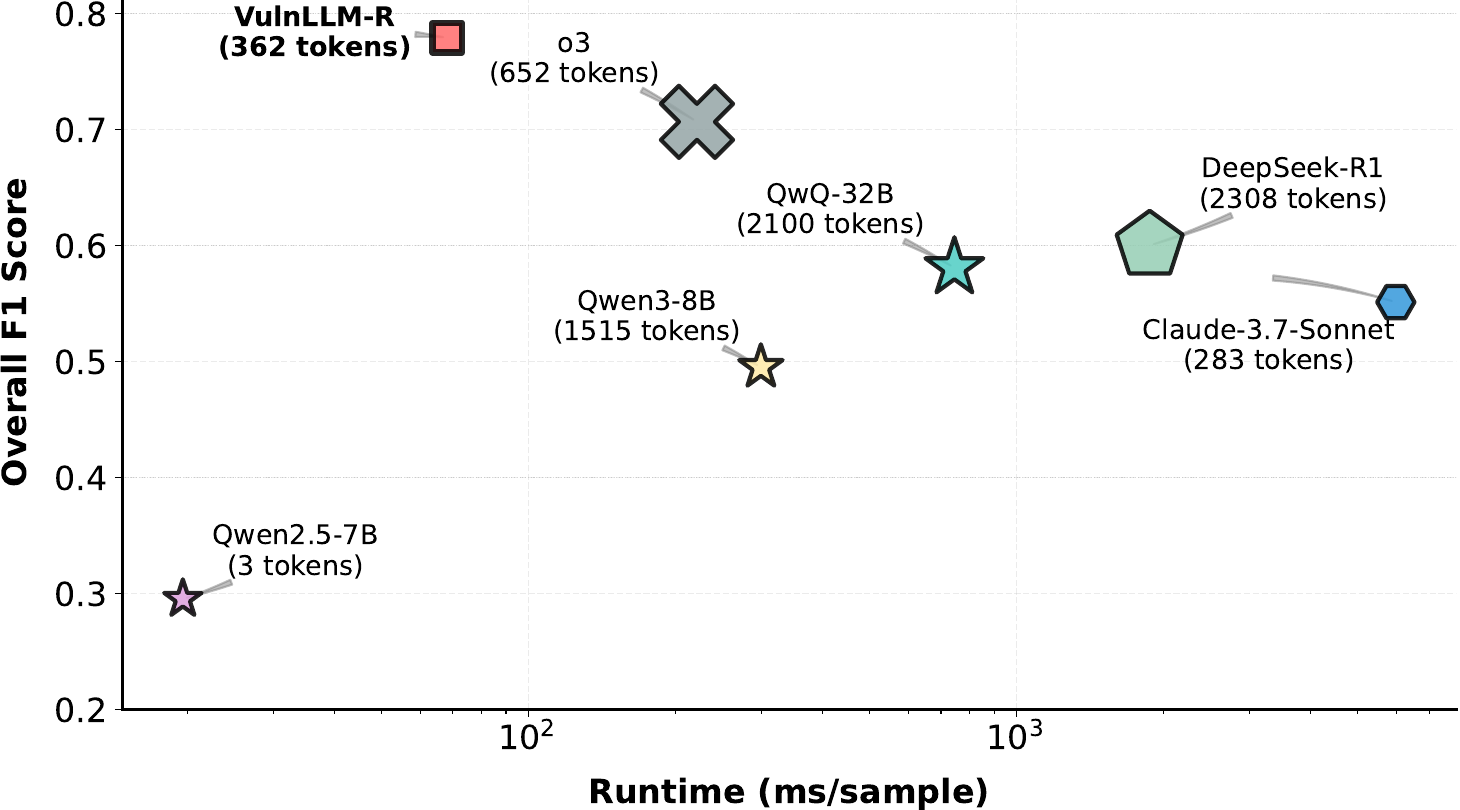}
  \caption{Average runtime vs. F1 score comparison between \sys and selected baselines on all benchmarks.}
  \label{fig:runtime_comparison}
\end{figure}

More importantly, \Cref{fig:main} further shows that \sys outperforms all baseline LLMs. 
First, \sys beats its teacher models \dsr and \qwq on all tasks.
This is \emph{non-trivial} as the teacher model is typically the upper bound of the student model. 
We believe \sys can push its limit because of our strategies of having two teacher models, as well as our reasoning filtering and correction mechanism.
These strategies can reduce the effect of individual teacher models' errors to \sys.
The constitution-based correction further enables \sys to learn new knowledge that is not acquired by its teacher models. 
Our unique training recipe further pushes \sys to outperform SOTA commercial models.
This validates our argument in~\Cref{sec:tech_overview} that vulnerability detection is a unique task that requires specialized models.
This is a remarkable achievement as \sys is at least 30$\times$ smaller than these commercial models. 
This is also the first time~\textit{to show that specialized small models can beat SOTA commercial models on security-specific tasks.}
Furthermore, \sys achieves the highest performance on complex datasets (Juliet: multi-functions, PrimeVul: long functions, Arvo: multi-functions), demonstrating its capabilities of analyzing long functions and vulnerabilities that involve multiple functions.
\sys can also generalize to unseen CWEs, as shown in~\Cref{app:ood_baseline}, which confirms its practicality. 
Finally, we surprisingly find that~\emph{although not being trained with Java}, \sys still outperforms SOTA models on the Java datasets, further~\emph{demonstrating its generalizability}.
\Cref{fig:runtime_comparison} further demonstrates the advantage of \sys over SOTA models in efficiency.
The superior speed of \sys is attributed to two key factors: smaller models and summary-based training (shortening average output tokens and reducing inference time).

\noindent\textbf{Error Analysis.}
We find that \sys often makes false positives because of the lack of context.
For example, \sys often suspects the input to belong to CWE-367 (Race Condition) even though the input is benign.
The race condition is a vulnerability that typically requires running the program to determine.
Without the dynamic information, LLMs tend to be conservative and output this vulnerability if there may be multiple processes involved.
Additionally, we observe false negatives stemming from data quality issues, particularly in \primevul, where benign code is sometimes mislabeled as vulnerable.
Please see examples and detailed analysis in~\Cref{app:error_analysis}.

\subsection{Ablation Study}
\label{sec:eval_ablation}

\begin{figure}
  \centering
  \includegraphics[width=0.8\linewidth]{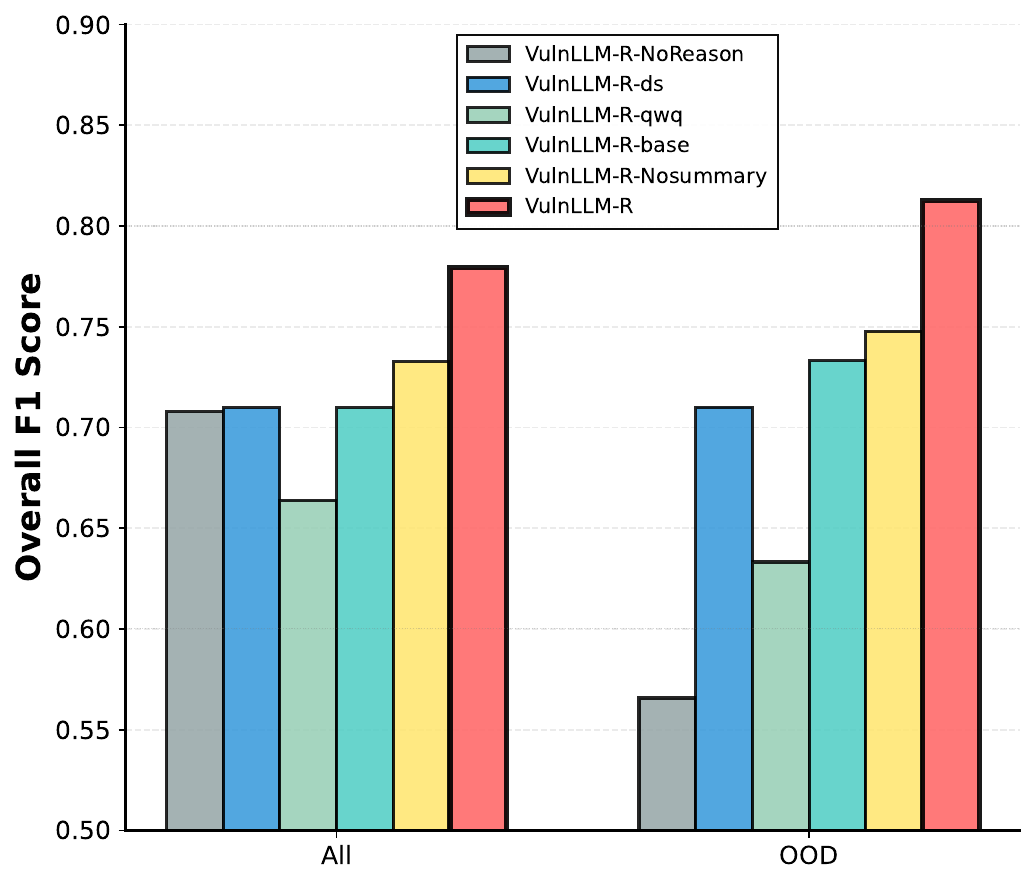}
  \caption{Comparison results between \sys and \sys-NoReason on all testing data.}
  \label{fig:ablation}
\end{figure}

\noindent\textbf{\sys without reasoning.}
We first demonstrate the necessity of using a reasoning model.
Here, we train the \sys-NoReason model, where we directly fine-tune our base model to output the correct answer without the intermediate reasoning process. 
\Cref{fig:ablation} shows that \sys performs better than \sys-NoReason.
Especially, the result shows the advantage of \sys over \sys-NoReason in OOD CWEs, confirming the superior generalizability of reasoning models.
The result is aligned with our statement in~\Cref{sec:tech_overview} that the reasoning enables LLMs to handle more complex tasks and have better generalizability.

\noindent\textbf{\sys's reasoning data generation.}
Here, we validate the necessity of having two teacher models.
We train three models, using \dsr and \qwq (\sys-base), \dsr solely (\sys-ds), and \qwq (\sys-qwq) solely as teacher models.
Here, we directly use the raw reasoning data to train the model without applying our reasoning data filtering and summary-based fine-tuning.
\Cref{fig:ablation} shows that our model trained from two teacher models outperforms the other two models, validating the effectiveness of this design.
As discussed in~\Cref{sec:tech_overview}, \dsr and \qwq have different reasoning logic and structures. \dsr thinks at a higher level than \qwq (better at inferring program states).
As such, using both models as teachers can help our model learn from thinking structures without biasing towards one pattern that may cause model degeneration.
It also helps our model to balance the false positives and false negatives. 

\noindent\textbf{\sys's reasoning data filtering and constitution.}
Here, we train a model without the final summary-based training (\sys-NoSummary).
That is, we apply our proposed filtering and constitution-based correction method to post-process raw reasoning data and train the base model accordingly. 
Compared to \sys-base, \sys-NoSummary has the additional step of data filtering and constitution-based correction.
As shown in~\Cref{fig:ablation}, \sys-NoSummary shows better performance than \sys-base across all datasets.
This result confirms the effectiveness of our unique filtering and correction mechanism. 
As discussed in~\Cref{sec:tech_overview}, this is different from existing distillation recipes for general math and coding tasks. 
We believe having data with correct answers is important for the model to learn the correct knowledge that is lacking in the base model. 

\noindent\textbf{\sys's summary-based fine-tuning.}
Finally, \Cref{fig:ablation} compares \sys-NoSummary and \sys and validates the effectiveness of our summary-based fine-tuning.
It also improves the testing-phase efficiency of our model by 80\% as this process significantly shortens the reasoning length of our model.
We believe the performance improvement is because, after the summary-based fine-tuning, our model has fewer overthinking cases.
On the one hand, this protects the model from going beyond its context limit; on the other hand, it also prevents the model from degeneration (overthinking may lead to degeneration as the model keeps outputting similar reasoning).

\subsection{\sys under testing-phase scaling}
\label{sec:eval_testing}

\noindent\textbf{Setup and design.}
In this experiment, we evaluate whether testing-phase scaling is helpful for our task, specifically focusing on token usage analysis and parallel scaling. 
For token usage analysis, we calculate the average number of tokens generated by different models under a maximum length constraint of 8,192 tokens to understand their generation characteristics and efficiency.
For parallel scaling, we fix the maximum output length to 8,192 tokens per generation and explore the impact of multiple parallel queries by setting the number of queries to 1, 4, 8, and 16.
For each configuration, we generate multiple responses and aggregate the final results through majority voting.
We report the performance of our model under the parallel scaling strategy and provide an analysis of average token usage patterns across the selected datasets to assess their respective effectiveness.
\begin{figure}
  \centering
  \includegraphics[width=0.63\linewidth]{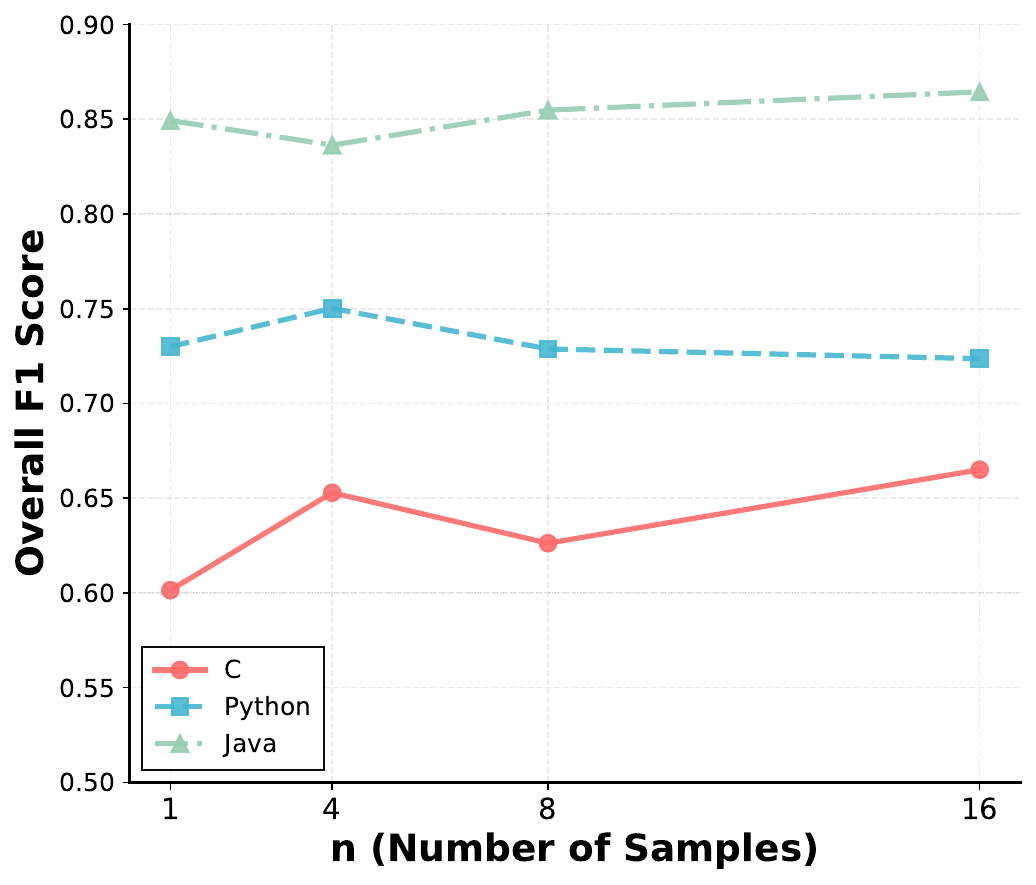}
  \caption{Impact of parallel scaling on vulnerability detection performance.
  We evaluate \sys with varying numbers of parallel queries (1, 4, 8, 16).}
  \label{fig:testing}
\end{figure}

\noindent\textbf{Results.}
As shown in~\Cref{fig:runtime_comparison}, our model achieves remarkable efficiency in token usage.
While the non-reasoning model \qweninstruct produces the shortest outputs, our model generates notably concise responses with an average of 362 tokens, approaching another non-reasoning model \claude (283 tokens).
This demonstrates our model's efficiency in providing concise yet accurate vulnerability detection, primarily due to our efficiency training.
Our results in~\Cref{fig:testing} demonstrate that increasing the number of parallel queries does not consistently improve vulnerability detection performance through majority voting, with minimal or negligible gains observed beyond 4-16 queries.
This differs from the observation in general coding and math tasks, where long reasoning or multiple sampling can improve model performance~\cite{muennighoff2025s1}.
This unique observation shows that for vulnerability detection, the model can typically reach the correct conclusion within a short output length and with a moderate number of parallel samples.

\section{\sys Agent in Real-world Projects}
\label{sec:application}

\subsection{Agent Scaffold}

We implement an agent scaffold consisting of two components: function selector and context retriever.

\noindent\textbf{Function selector.}
The function selector identifies target functions for \sys to analyze for vulnerabilities.
First, we construct a call graph that contains direct calls discovered by \codeql, and augmented with potential indirect calls identified via type‐based analysis~\cite{diwan1998type}.
We choose type‐based analysis for its speed and efficiency.
Specifically, to identify potential indirect calls, we collect all functions whose addresses are assigned to function pointers. We then identify all call sites that invoke functions through function pointers, and infer an indirect call edge when both the number and types of parameters match exactly.
Finally, any function reachable via a call graph path from any harness entry point is marked as a target function.

\noindent\textbf{Context retriever.}
The context retriever gathers relevant context for each target function. To achieve this, we adopt a hybrid strategy that combines rule-based static collection for the initial context with agent-based dynamic retrieval when additional information is needed.

First, for each target function, we sample three random call paths from each harness entry point to the target function, and include the implementation of every intermediate function along these paths as the initial context. This design allows \sys to reason how the target function is invoked from the entry point, which is often critical for reasoning about potential vulnerable conditions. The choice of three paths strikes a balance between preserving contextual completeness and staying within \sys's input length constraints.

Second, the model is equipped with a tool capable of retrieving the implementation of any function by name, enabling it to autonomously obtain missing context during analysis. This mechanism addresses cases where relevant functions may not appear along the sampled call paths but can still influence the target function—either by setting global variables, modifying parameters passed to it, or being among its callees whose behavior is essential for vulnerability reasoning. Because such potentially relevant context is too diverse to be captured precisely through static rules without introducing significant noise, we let \sys selectively fetch task-specific information through LLM-based tool calls.

\noindent\textbf{Agentic training.}
To strengthen multi-step reasoning and effective tool use, we train \sys using trajectories collected with the scaffold above.
A similar approach has been adopted in recently released models with agentic capabilities, such as Kimi-K2~\cite{kimik2} and Qwen3-Coder-480B-A35B-Instruct~\cite{Qwen3Coder}.
As described in~\Cref{sec:tech}, for each project in the training set we run the agent and record every query and response, details of available tools, the model’s tool call actions, and the corresponding tool outputs throughout execution.
Using the collected trajectories, we then perform standard SFT, where prompt tokens—including user queries, system instructions, historical conversation, and tool call execution results—are masked during training. 
Only the teacher model’s generated outputs, which comprise reasoning steps and explicit tool call actions such as the arguments and invocation format, are used as the prediction targets for our model.
This training improves our model’s ability to identify functions that are critical for vulnerability reasoning yet missing from the current context, and enables the model to retrieve their implementations via tool calls.

\subsection{Experiment Design and Results}

\noindent\textbf{Benchmark selection.}
In order to evaluate our project-level vulnerability detection capabilities, we select five projects provided by DARPA for its AI Cyber Challenge (AIxCC) competition, which are deliberately modified versions of widely used real-world projects. 
These projects are Nginx~\cite{nginx2025}, FreeRDP~\cite{freerdp}, libexif~\cite{libexif}, Tika~\cite{tika} and ZooKeeper~\cite{zookeeper}.
We choose the popular Nginx\cite{nginx2025} along with four other projects that are released after the latest model cutoff, together covering both C and Java. 
The details of each target project are described in ~\Cref{app:agent_data_targets}.
During fine‐tuning, our system’s training set did not include any data from the selected projects.

For the AIxCC C/C++ targets, seven CWEs are in scope: CWE-125, CWE-787, CWE-119, CWE-416, CWE-415, CWE-476, and CWE-190. 
Because CWE-119 is the parent of both CWE-125 and CWE-787, and CWE-416 and CWE-415 can be interchanged, we consolidate them to avoid ambiguity. 
In our evaluation, CWE-125 and CWE-787 are mapped to CWE-119, CWE-415 is mapped to CWE-416, and this results in four unique CWEs in scope: CWE-119, CWE-190, CWE-416, and CWE-476.
For Java targets, 13 CWES are in scope, covering CWE-74, CWE-22, CWE-918, CWE-502, CWE-917, CWE-90, CWE-154, CWE-470, CWE-777, CWE-89, CWE-643, CWE-611, and CWE-835.

\noindent\textbf{Baseline.} 
We compare \sys against SOTA traditional vulnerability detection tools (both static and dynamic) as well as leading commercial models. 
For static analysis, we select the popular static analysis tool \codeql v2.21.3 (the latest release). 
We run both the default and the security‑extended query sets on each target project.
For dynamic analysis, we choose \aflpp v4.32c (the latest release) for C/C++ targets, because it consistently achieves top performance in fuzzing benchmarks such as FuzzBench~\cite{metzman2021fuzzbench}. 
For Java targets, we choose the widely-used fuzzer \jazzer.
In addition, we include LLM-assisted methods as additional baselines for both static and dynamic analysis. 
For static analysis, we adopt \repoaudit~\cite{guo2025repoaudit}, an LLM-augmented static analysis framework. 
For dynamic analysis, we adopt \gtwofuzz~\cite{zhang2025low} for the C/C++ targets, which leverages an LLM for seed generation. 
In addition to these open‐source baselines, we also evaluate two leading commercial LLMs (i.e., \othree and \claude).
For each fuzzer (i.e., \aflpp, \jazzer and \gtwofuzz), we execute one fuzzer instance per harness for 24 hours, pinned to a single CPU core.
For all evaluated LLMs, we employ our agent scaffold for context collection. 
\sys itself is served through \texttt{vLLM 0.7.2} on one NVIDIA H100 GPU.
To enable a fair comparison with fuzzers, we consider a vulnerability as detected by static analysis tools and LLMs if at least one of its corresponding vulnerable functions, i.e., the functions modified in the golden patch provided by the AIxCC organizers, is deemed vulnerable.

\label{sec:project_level_res}
\begin{figure}[t]
    \centering
    \includegraphics[width=1\linewidth]{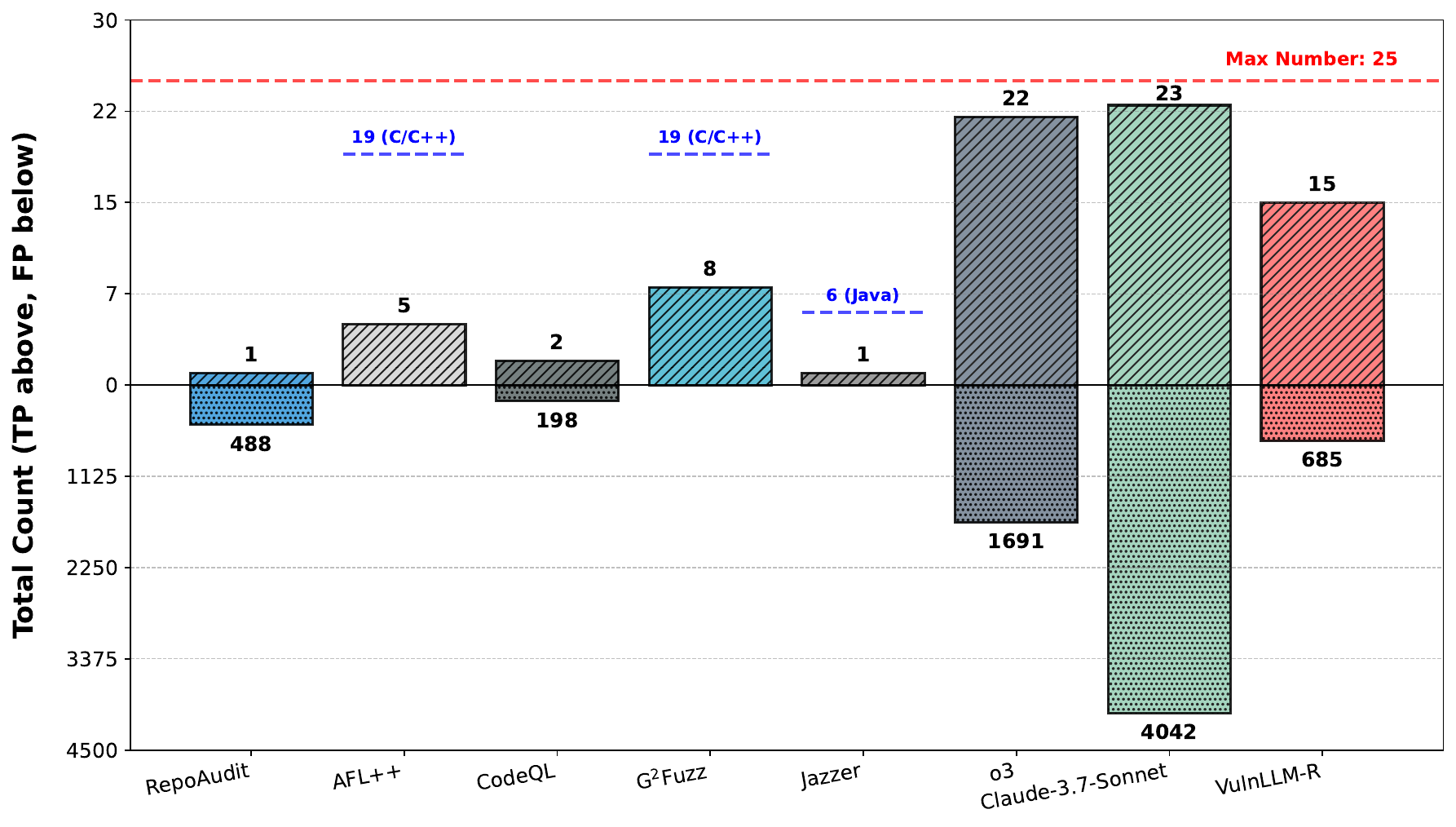}
    \caption{Project-level TP/FP comparison of \sys and all baselines across all targets.}
    \label{fig:project_level_res}
\end{figure}

\noindent\textbf{Experiment results.}
~\Cref{fig:project_level_res} shows the comparative results of \sys and all baselines. 
In particular, agentic-based methods (agent + \sys, agent + \claude and agent + \othree) consistently detected the largest number of vulnerabilities across all five targets. 

In contrast, static-analysis methods, including \codeql and \repoaudit, identified only a small fraction of vulnerabilities due to their reliance on static rule patterns and their limited support for diverse vulnerability types. 
Although \repoaudit benefits from LLM assistance, its core detection mechanism still follows a taint-analysis design, and it also relies on pattern matching to extract sources and sinks.
In addition, \repoaudit tracks data-flow facts across no more than four functions, a restriction that further limits its ability to capture complex, multi-function vulnerabilities.
As a detailed result shown in~\Cref{fig:all_project_level_res}, \codeql detected only two vulnerabilities (both in Tika), while \repoaudit identified just a single vulnerability (in Nginx).
 
For traditional and LLM-assisted fuzzers, despite inherently avoiding false positives due to the nature of fuzzing, they discover relatively few vulnerabilities and require substantial time (24 hours per harness). 
By contrast, \sys complete detection on each of the five targets within one hour. 
Moreover, fuzzers provide limited insights into which functions are vulnerable and why; significant additional triage and debugging effort is required to understand the root causes. 

For commercial LLMs, we observe that \claude tends to produce a large number of false positives.
For example, in FreeRDP, \claude predicts 1834 target functions out of 3347 as vulnerable, which is clearly unacceptable for practical use.
\othree exhibits a similar issue on FreeRDP, where 1057 functions are incorrectly classified as vulnerable.
Despite having only 7B parameters, \sys achieves comparable performance with leading commercial models in terms of recall, consistently outperforming non-agentic baselines, and yields the lowest false-positive rate among all agentic baselines with commercial models.
The gap in recall between \sys and commercial models mainly appears on Nginx, where our method sometimes fails to capture interprocedural interactions across distant call-graph nodes (typical for use-after-free cases, e.g., the registration and later processing of request headers). 
Surprisingly, \othree and \claude can still make correct predictions in such cases even without explicit contextual information, e.g., they appear to know Nginx's request-header processing mechanism even though this information is never provided as input, suggesting that \othree and \claude may rely on prior knowledge of Nginx internals—potentially due to data leakage.



\section{Discussion}
\label{sec:discussion}

\noindent\textbf{More languages and CWEs.}
In this paper, we focus on the most widely used programming languages (Python, C/C++, Java) and the most critical CWEs.
Future works can extend \sys to more languages and vulnerability types by collecting new datasets and developing more advanced training recipes. 
First, we lack high-quality benchmarks for broader CWEs and languages. 
Specifically, although existing benchmarks cover broader CWEs, the scale and quality are limited.
For example, while \primevul contains over 100 CWEs, many have only a few data points and contain label errors. 
Similarly, most existing vulnerability datasets are about Python, C/C++, and Java~\cite{sheng2025llmssoftwaresecuritysurvey}.
Other widely used languages, such as Go, TypeScript, and Rust, lack large-scale and high-quality datasets with language-specific vulnerabilities. 
As such, we call for the community's collaborative efforts to enrich the landscape of vulnerability detection benchmarks.  
Second, it is non-trivial to train one LLM to handle multiple languages, as it may require deliberate training data blending strategies~\cite{chen2024rise}, which we defer to future work.

\noindent\textbf{Project-level vulnerability detection.}
Existing ML-based vulnerability detection mainly stays at the individual function level.
In this work, we show promising results of conducting project-level vulnerability detection with reasoning model-driven agents.
Future efforts can explore three directions.
First, we are significantly short of project-level benchmarks with real-world projects.
\arvo and OSS-Fuzz~\cite{ossfuzz} can serve as a solid starting point for constructing such benchmarks.
Second, we need to construct more advanced agents that integrate richer tool sets for context retrieval and supporting analysis. 
Third, we can further train specialized LLMs that either conduct tool call scheduling~\cite{SkyRL} or serve as individual components in the agentic system~\cite{tang2025co}.

\noindent\textbf{Reasoning model to LLM agents.}
On the model side, future work can further improve the quality of the models' reasoning chains and further eliminate potential shortcuts.  
For agents, more advanced tools will further improve AI's capability in analyzing complex projects. 
For example, we can integrate LLMs with more static analysis tools (e.g., RepoAudit~\cite{guo2025repoaudit}).
We can further integrate RAG in the agent system or even integrate dynamic fuzzers to reduce false positives. 
Our trained model can serve as an important component in such agentic systems to identify potentially vulnerable code.

\section{Related Work}
\label{sec:related-work}

\noindent\textbf{Static analysis} detects potential vulnerabilities in source code without executing it~\cite{ye2023viper,li2024lrminer}.
Most existing static analysis methods and tools conduct inter-procedural analysis, which examines interactions across the entire program. 
Certain static analysis tools have been developed and widely adopted in real-world vulnerability detection, including CodeQL~\cite{codeql}, Bandit~\cite{bandit}, LlamaFireWall~\cite{llamafirewall}, and Infer~\cite{infer}. 
At a high level, these tools integrate predefined heuristics and rules to identify vulnerabilities through pattern-matching techniques.
However, static analysis approaches face several fundamental limitations. 
First, they are neither complete nor sound: they suffer from false positive issues due to overly conservative analysis and false negatives due to the limited generalizability of their rule-based approaches. 
Additionally, the process of developing and maintaining comprehensive rule sets is both time-consuming and requires substantial domain expertise, making it difficult to scale across different programming languages, frameworks, and evolving vulnerability patterns.




\noindent\textbf{Fuzzing} is a widely used dynamic analysis method for vulnerability discovery, which continuously generates new inputs through mutation.
The primary goal of fuzzing is to improve program coverage and thus increase the likelihood of triggering bugs. 
The SOTA fuzzer is \aflpp~\cite{AFLplusplusWoot20}, with numerous improvements built upon it to enhance seed generation~\cite{yue2020ecofuzz, wang2021reinforcement, zhao2021evolutionary, lyu2022slime}, power scheduling~\cite{bohme2016CoveragebasedGreyboxFuzzing,yue2020ecofuzz}, and mutator designs~\cite{koike2022slopt,wu2022one,wang2021cmfuzz,karamcheti2018adaptive,scott2021banditfuzz}.
More recently, researchers leverage LLMs to improve fuzzing effectiveness~\cite{meng2024large,deng2023large,xia2024fuzz4all,asmita2024fuzzing,lyu2024prompt}. 
Despite these research efforts, \aflpp~\cite{AFLplusplusWoot20} remains the most widely adopted fuzzer in practice.
Fuzzing offers several advantages: it does not introduce false positives and can generate actual vulnerability-triggering inputs. 
Dynamic analysis through fuzzing can be particularly effective in detecting certain types of vulnerabilities, such as buffer overflows, as it can trigger memory corruption and crash conditions in real-time.
However, fuzzing faces significant limitations in terms of efficiency. 
Due to its reliance on random mutation strategies, fuzzing typically achieves low code coverage and struggles to reach deep program branches, limiting its ability to discover complex vulnerabilities. 
Note that there are some techniques for directed fuzzing~\cite{bohme2017directed} and domain-specific fuzzing~\cite{guo2023operand}, these techniques are not aligned with our problem scope.


\noindent\textbf{Deep learning and LLMs.}
In recent years, deep learning-based methods are increasingly applied to vulnerability detection tasks.
Traditional deep learning approaches use graph neural networks~\cite{zhou2019devign,xu2017neural,steenhoek2024dataflow,GHAFFARIAN2021189,xu2019vulsniper}, recurrent neural networks~\cite{Zou2019vul,wang2023deepvd}, and convolutional models~\cite{le2022deepcva,wu2022vulcnn} to train vulnerability detection models.
However, these approaches face several limitations: they are typically small models that cannot handle complex functions, let alone entire projects; they generalize poorly across different languages and CWE categories; and they are prone to learning shortcuts~\cite{steenhoek2025errmachine}, where models make decisions based on superficial code signatures rather than reasoning about actual vulnerabilities.

More recently, researchers explore the use of large language models for vulnerability detection, which can be broadly categorized into three approaches: fine-tuning existing LLMs~\cite{ding2024primevul,du-etal-2024-generalization,he2023large,yang2024large,shestov2024finetuning}, Retrieval Augmented Generation (RAG) techniques~\cite{du2024vulragenhancing}, and LLM-facilitated static analysis.
Fine-tuning approaches, while showing promise, still rely on relatively small models without reasoning capabilities, thus suffering from similar limitations as traditional deep learning methods. 
RAG-based approaches incorporate external knowledge sources, such as CWE databases, to enhance detection capabilities, which can be integrated with our proposed method to further improve performance.
Finally, recent work~\cite{li2024iris,guo2025repoaudit} uses LLMs to facilitate static taint analysis by analyzing data flow or identifying potential sinks or sources, which is less flexible than our agentic-based solution.
LLMxCPG~\cite{lekssays2025llmxcpg} represents another approach that combines code property graphs with LLMs for vulnerability detection.
It employs Joern-based taint analysis to extract context from source code and uses this extracted information to guide the LLM's analysis.
However, LLMxCPG operates as a pure classification model that outputs a single-word prediction, lacking the reasoning capabilities that enable deeper understanding of complex vulnerabilities.

\section{Conclusion}

We introduce \sys, the first specialized reasoning model for vulnerability detection. 
We propose a novel training recipe with data selection, reasoning data generation, reasoning data filtering and correction, and testing-phase optimization. 
We train a 7B model and demonstrate its superiority over SOTA static analysis tools, and large open-source and commercial models in detecting vulnerable functions.
We construct an agent based on our model and show that it outperforms SOTA static and dynamic analysis tools in 5 real-world projects.
With extensive evaluation, we can conclude that vulnerability detection can benefit from specialized small reasoning models as well as well-crafted agents.

\bibliographystyle{icml2025}
\bibliography{conferences,ref}

\appendix

\section{More Details on Dataset Construction}

\subsection{Sample Selection}
\label{app:sample_selection}
In this section, we introduce the criteria for data selection from \primevul, \arvo, and \juliet datasets.

\noindent\textbf{\primevul.}
In \primevul dataset, we manually check some samples and find that many of them lack sufficient context, making it difficult to judge the presence of a vulnerability reliably. 

In this case, we adopt an LLM-assisted human verification pipeline.
Specifically, we manually analyze each sample to assess whether the provided context is sufficient. 
The specific prompt is provided below.
\begin{tcolorbox}[colback=white,title={Context Assessment Prompt}, colframe=black, breakable]
You are an advanced vulnerability detection model. 
Your task is to analyze whether the provided context is sufficient to make a vulnerability determination in the target function. Specifically, you need to check whether the provided context contains enough information to understand the data flow, security boundaries, and potential attack vectors.

You are given the following context and code snippet:

Context:
\begin{verbatim}
{context}
\end{verbatim}

Target Function:
\begin{verbatim}
{CODE}
\end{verbatim}

You should STRICTLY structure your response as follows:

\begin{verbatim}
## Final Answer
#judge: <yes/no>
#function: list of needed functions
\end{verbatim}

Examples:

If the context is sufficient:
\begin{verbatim}
## Final Answer
#judge: yes
#function: N/A
\end{verbatim}
If the context is insufficient:
\begin{verbatim}
## Final Answer
#judge: no
#function: [a, b, c]
\end{verbatim}
    

\end{tcolorbox}
Using the prompt above, we query \othree, \claude, and \dsr twice for each sample to determine whether the corresponding context is sufficient.
If any model indicates insufficient context in either of the two queries, the sample undergoes human verification. 
Samples judged to have insufficient context are then filtered out.

\noindent\textbf{\arvo.}
According to \arvo dataset construction in~\Cref{sec:data_collection}, we need to collect three components: (1) vulnerable code; (2) patched code; (3) necessary context. 
In this process, we need to verify if (1) vulnerable version trigger the crash, (2) patched version does not trigger the crash.
Besides, to simplify the data collection, we filter out samples that the corresponding patch is a part of a large commit that includes many files, or the patch is across multiple commits.
By doing this, we collect 1399 samples in total, which are then split into training and testing sets.

\noindent\textbf{\juliet.}
For each CWE in the \juliet dataset, we rank all available C/C++ test case files by their length in lines of code and select the longest files.
For each CWE, up to 10 files are selected for the testing set and up to 100 files for the training set.
Each file is then split into one vulnerable and one benign sample.
If fewer than 110 files are available, 10 files are first assigned to the testing set, and the remaining files (up to 100) to the training set. 

\subsection{Data Pre-Process}
\label{app:preprocess_data}

In this section, we describe in detail the pre-processing steps that are applied to each open-source dataset in~\Cref{tab:dataset-composition}.
We explain in detail how we construct the paired samples of vulnerable and benign code from \juliet and \arvo.
For all other datasets, we do not need to perform extra pre-processing.

\noindent\textbf{\juliet.}
The Juliet data often has a strongly entangled vulnerable and patched code, i.e., the vulnerable and patched functions are in the same file or even the same class. 
Besides, the code typically has obvious indicators in the function names (patched and vulnerable code are named as `good()' and `bad()').
Here, to construct a paired code sample, we use slightly different strategies for C/C++ and Java.

We process each sample in \juliet C/C++ dataset to construct paired samples of vulnerable and benign code. The overall pipeline consists of the following steps:

\textit{Macro Block Processing.}
Each \juliet C/C++ dataset sample typically contains both vulnerable and benign code implementations, separated by preprocessor macros such as \#ifndef OMITBAD (for vulnerable code) and \#ifndef OMITGOOD (for benign code).
To obtain the benign version, we remove all code blocks guarded by \#ifndef OMITBAD (i.e., delete both the macro lines and the inner content), and only strip the macro lines for \#ifndef OMITGOOD, keeping the content inside.
To obtain the vulnerable version, we remove all code blocks guarded by \#ifndef OMITGOOD, and only strip the macro lines for \#ifndef OMITBAD, again preserving the inner content.

\textit{Code Cleaning and Obfuscation.}
To avoid information leakage, We remove all comments.
Any function names containing keywords like "good" or "bad" are replaced with random names.
Other easily recognizable keywords (such as ``CWE'' identifiers, or good/bad in print statements or variables) are deleted from the code.
Namespace declarations and usages are also replaced with a fixed string to prevent leaking information via namespace names.

For \juliet Java dataset, we first use JavaParser~\cite{java_parser} to identify the good functions based on their function names.
We delete the identified good functions from the same file, leaving the rest as a vulnerable sample. 
To remove the obvious indicators in the code, we design a comprehensive obfuscation strategy, including removing all source code comments, removing package declarations at the beginning of the files, and obfuscating identifiers.
Any class names, method names, or variable names containing keywords like "cwe", "good", "bad", "G2B", as well as string literals in output statements, are replaced with unique, randomly generated 7-character strings. 

\noindent\textbf{\arvo.}
In \arvo dataset, the ground-truth CWE label is not given, so we infer the CWE number from the vulnerable program crash, specifically, our matching rules are shown in~\Cref{tab:arvo-cwe-matching}.


\begin{table}[ht]
    \centering
    \caption{CWE Classification Rules and Detection Patterns}
    \label{tab:arvo-cwe-matching}
    \resizebox{0.45\textwidth}{!}{
    \begin{tabular}{l|p{4cm}}
    \hline
    \textbf{CWE Number} & \textbf{Detection Rules} \\
    \hline
    \multirow{1}{*}{\makecell[l]{\textbf{CWE-125}\\Out of Bound Read}} & 
    Contains "overflow" in sanitizer output AND "read" in sanitizer output \\
    \hline
    \multirow{1}{*}{\makecell[l]{\textbf{CWE-416}\\Use After Free}} & 
    Contains "use" and "free" in sanitizer output \\
    \hline
    \multirow{1}{*}{\makecell[l]{\textbf{CWE-787}\\Out of Bound Write}} & 
    Contains "overflow" AND "write" in sanitizer output \\
    \hline
    \end{tabular}}
\end{table}

\section{Additional Experiments and Results on Function-level Evaluation}

\subsection{Out-of-Distribution CWEs Results}
\label{app:ood_baseline}

We evaluate the generalization capability of \sys on OOD CWEs and unseen programming languages. 
As shown in~\Cref{tab:ood_cwe_results}, \sys demonstrates strong generalization performance across different settings. 
For Python, \sys achieves nearly identical performance on OOD CWEs (0.719) compared to the full dataset (0.723), with only a negligible 0.6\% decrease. 
Similarly, for C, \sys even shows a 5.29\% improvement on OOD CWEs (0.776 vs. 0.737). 

Moreover, we evaluate \sys on Java, which represents a completely unseen language not included in the training data. 
Remarkably, \sys achieves an F1 score of 0.870 on Java, demonstrating excellent cross-language generalization capabilities. 
In contrast, the ablated models show significant performance degradation on OOD data. 
\sys-NoReason suffers the most severe drops, with 37.6\% decrease in Python and 12.9\% in C, achieving only 0.727 on Java. 
\sys-base also exhibits substantial performance drops of 14.1\% in Python and 36.9\% in C.
These results show that the reasoning training not only helps the model understand vulnerability patterns more deeply but also enables it to transfer this understanding to unseen CWEs and new programming languages.

\begin{table}[ht]
    \centering
    \caption{OOD CWE results for \sys across languages}
    \label{tab:ood_cwe_results}
    \setlength{\tabcolsep}{4pt}
    \resizebox{\linewidth}{!}{
    \begin{tabular}{@{} l  c  c  c  c c @{}}
      \toprule
      \multicolumn{1}{c}{\textbf{}} 
        & \multicolumn{2}{c}{\textbf{Python}} 
        & \multicolumn{2}{c}{\textbf{C}} & \multicolumn{1}{c}{\textbf{Java}}\\
      \cmidrule(lr){2-3} \cmidrule(lr){4-5} \cmidrule(lr){6-6}
        & \textbf{All} & \textbf{OOD} 
        & \textbf{All} & \textbf{OOD} & \textbf{OOD}\\
     \sys 
        & 0.723 & 0.719 (-0.6\%) & 0.737 & 0.776 (+5.29\%) & 0.870\\
      \sys-NoReason
        & 0.797 & 0.497 (-37.6\%) & 0.784 & 0.683 (-12.9\%) & 0.727\\
        \sys-base
        & 0.675 & 0.580 (-14.1\%) & 0.570 & 0.359 (-36.9\%) & 0.690\\
      \bottomrule
    \end{tabular}
    }
\end{table}

\subsection{Error Analysis}
\label{app:error_analysis}
\subsubsection{False Positive in TOCTOU Vulnerability Detection}

\noindent\textbf{Code Snippet}

\begin{lstlisting}[language=C, basicstyle=\listingsttfamily, breaklines=true, columns=flexible, keepspaces=true]
set_interface_var(const char *iface,
                  const char var, const char name,
                  uint32_t val)
{
    FILE *fp;
    char spath[64+IFNAMSIZ];    /* XXX: magic constant */
    if (snprintf(spath, sizeof(spath), var, iface) >= sizeof(spath))
        return -1;
    if (access(spath, F_OK) != 0)
        return -1;
    fp = fopen(spath, "w");
    if (!fp) {
        if (name)
            flog(LOG_ERR, "failed to set %s (%u) for %s: %s",
                 name, val, iface, strerror(errno));
        return -1;
    }
    fprintf(fp, "%u", val);
    fclose(fp);
    return 0;
}
\end{lstlisting}

\noindent\textbf{Model Analysis}
\begin{quote}
The critical issue arises after the access check. The code verifies the file's existence but doesn't ensure the file is opened with proper permissions. If the file exists and is writable by the process, it proceeds to write. However, an attacker could replace the file with a symlink after the access check but before fopen. This is a Time-of-Check to Time-of-Use (TOCTOU) race condition (CWE-367), allowing unintended file writes.
\end{quote}

\noindent\textbf{Error Explanation}
The analysis incorrectly identifies a CWE-367 (TOCTOU) vulnerability due to \textbf{missing contextual information}. The sequential use of \texttt{access()} and \texttt{fopen()} appears vulnerable to race conditions where an attacker could substitute a symbolic link between these calls. However, this assessment lacks crucial context:
\begin{itemize}
    \item The \texttt{var} parameter likely contains format strings for system interface paths (e.g., \texttt{/proc/sys/} or \texttt{/sys/}), which are protected kernel directories where unprivileged users cannot create symbolic links
    \item The execution environment and permission model prevent the assumed attack vector
\end{itemize}

\subsubsection{Incomplete Patch Representation in \primevul}

\noindent\textbf{Code Snippet (Vulnerable)}

\begin{lstlisting}[language=C, basicstyle=\listingsttfamily, breaklines=true, columns=flexible, keepspaces=true]
static inline LineContribType * _gdContributionsAlloc(
    unsigned int line_length, unsigned int windows_size)
{
    unsigned int u = 0;
    LineContribType *res;
    // ... allocation code ...
    for (u = 0 ; u < line_length ; u++) {
        if (overflow2(windows_size, sizeof(double))) {
            overflow_error = 1;
        } else {
            res->ContribRow[u].Weights = (double *) 
                gdMalloc(windows_size * sizeof(double));
        }
        if (overflow_error == 1 || res->ContribRow[u].Weights == NULL) {
            u--;
            while (u >= 0) {  // BUG: unsigned comparison always true
                gdFree(res->ContribRow[u].Weights);
                u--;
            }
            return NULL;
        }
    }
    return res;
}
\end{lstlisting}

\noindent\textbf{Code Snippet (Patched)}

\begin{lstlisting}[language=C, basicstyle=\listingsttfamily, breaklines=true, columns=flexible, keepspaces=true]
// ... same allocation code ...
if (overflow_error == 1 || res->ContribRow[u].Weights == NULL) {
    unsigned int i;
    u--;
    for (i=0; i<=u; i++) {  // Fixed: proper iteration
        gdFree(res->ContribRow[i].Weights);
    }
    gdFree(res);  // Added: free res structure
    return NULL;
}
\end{lstlisting}

\noindent\textbf{Data Issue.}
This case exemplifies a critical \textbf{dataset quality problem} where some of the \primevul samples misrepresents vulnerability patches. 
The dataset labels CVE-2016-6207 as fixed, but the provided patch represents an intermediate commit that itself introduced CVE-2016-10166.
The original vulnerability (integer underflow in \texttt{while (u >= 0)} causing out-of-bounds memory access when unsigned \texttt{u} wraps around from 0 to UINT\_MAX) is only partially addressed.

\subsection{Hyper-Parameters Experiments}
\label{app:hyper-parameter}
In this section, we analyze the impact of different hyper-parameters on the performance of \sys.
We vary the number of queries for policy generation, architecture of reasoning model, and the size of base model.

\begin{table}[ht]
    \centering
    \caption{Performance comparison across different model configurations}
    \label{tab:hyper-parameter-base}
    \setlength{\tabcolsep}{6pt}
    \begin{tabular}{@{} l  c  c  c @{}}
      \toprule
      \textbf{Model} & \textbf{Python} & \textbf{C/C++} & \textbf{Java}\\
      \midrule
      \sys & 0.730 & 0.601 & 0.849\\
      \sys-qwen3-8B & 0.798 & 0.649 & 0.837\\
      \sys-qwen2.5-32B & 0.754 & 0.683 & 0.851\\
      \bottomrule
    \end{tabular}
\end{table}

\Cref{tab:hyper-parameter-base} shows the performance comparison across different model configurations.
After switching from \qweninstruct to \qwenthree, despite the similar model parameters, we observe significant improvements, particularly in Python and C/C++.
We attribute this to the updated model architecture and the inclusion of more Python and C/C++ data in pre-training.
Furthermore, \qwenthree is inherently a reasoning model, which facilitates more effective knowledge absorption during training.
Meanwhile, Qwen2.5-32B-Instruct achieves higher results on C/C++.
We believe the C dataset is inherently more complex, and the 32B model has greater capability to handle complex tasks.

\section{More Details on Project-level Experiments}
\label{app:project-level}

\subsection{Details of Selected Projects}
\label{app:agent_data_targets}

\begin{table}[t]
\centering
\scriptsize
\caption{Details of selected projects in agentic training set.}
\label{tab:training_projects}
\begin{tabular*}{\linewidth}{@{\extracolsep{\fill}} l l l c}
\toprule
\textbf{Project} & \textbf{Lang.} & \textbf{CWE} & \textbf{Train Funcs} \\
\midrule
\multirow[t]{3}{*}{Assimp} & \multirow[t]{3}{*}{C++} & CWE-119 (17) & \multirow[t]{3}{*}{314} \\
                           &                         & CWE-416 (2)  &                          \\
                           &                         & CWE-476 (2)  &                          \\
\midrule
\multirow[t]{2}{*}{libxml2} & \multirow[t]{2}{*}{C}  & CWE-119 (1)  & \multirow[t]{2}{*}{100} \\
                            &                        & CWE-416 (1)  &                         \\
\midrule
SQLite3 & C & CWE-119 (2) & 300 \\
\midrule
\multirow[t]{2}{*}{CUPS} & \multirow[t]{2}{*}{C} & CWE-119 (3) & \multirow[t]{2}{*}{130} \\
                         &                       & CWE-476 (3) &                         \\
\midrule
zt-zip & Java & CWE-22 (2) & 139 \\
\midrule
\multirow[t]{4}{*}{Commons Compress} & \multirow[t]{4}{*}{Java} & CWE-918 (1) & \multirow[t]{4}{*}{200} \\
                                     &                         & CWE-74 (1)  &                          \\
                                     &                         & CWE-502 (1) &                          \\
                                     &                         & CWE-22 (1)  &                          \\
\bottomrule
\end{tabular*}
\end{table}

\begin{table}[t]
\centering
\scriptsize
\caption{Details of projects used for evaluation.}
\label{tab:testing_projects}
\begin{tabular*}{\linewidth}{@{\extracolsep{\fill}} l l l c}
\toprule
\textbf{Project} & \textbf{Lang.} & \textbf{CWE} & \textbf{Target Funcs} \\
\midrule
\multirow[t]{3}{*}{Nginx} & \multirow[t]{3}{*}{C} & CWE-119 (8)  & \multirow[t]{3}{*}{2076} \\
                          &                       & CWE-416 (4)  &                          \\
                          &                       & CWE-476 (2)  &                          \\
\midrule
\multirow[t]{2}{*}{FreeRDP} & \multirow[t]{2}{*}{C} & CWE-119 (1)  & \multirow[t]{2}{*}{3347} \\
                            &                       & CWE-476 (1)  &                          \\
\midrule
libexif   & C    & CWE-119 (3) & 225 \\
\midrule
\multirow[t]{4}{*}{Tika} & \multirow[t]{4}{*}{Java} & CWE-22 (1)   & \multirow[t]{4}{*}{2632} \\
                         &                         & CWE-74 (2)   &                          \\
                         &                         & CWE-918 (1)  &                          \\
                         &                         & CWE-611 (1)  &                          \\
\midrule
Zookeeper & Java & CWE-502 (1) & 546 \\
\bottomrule
\end{tabular*}
\end{table}

\Cref{tab:training_projects} summarizes the projects included in our agentic training set. 
For each project, we report its programming language, the CWEs observed in the ground-truth patches, 
and the number of functions selected into the training set.

\Cref{tab:testing_projects} lists the projects selected as evaluation targets, showing the corresponding CWE categories for each project and the number of target functions.

\subsection{Retrieved Context and Per-project Result}
\label{app:context}

We provide an example trace below to illustrate how the agent retrieves context and reasons about vulnerabilities. The agent calls the tool to retrieve the implementation of functions  
ngx\_pcalloc and ngx\_array\_init, and then uses this additional context to confirm that allocation, initialization, and pointer handling are safe, leading to the conclusion that no vulnerabilities are present.

\begin{tcolorbox}[colback=white,title={Example Trace}, colframe=black, breakable]

\textbf{User:}  
\begin{lstlisting}[numbers=none,breakautoindent=false,breakindent=0pt]
You are an advanced vulnerability detection model. Your task is to check if a specific vulnerability exists in a given piece of code.
...
You are given the following code snippet:  
// context: ...
// target function: 
static void *
ngx_core_module_create_conf(ngx_cycle_t *cycle)
{
...
}

You should STRICTLY structure your response as follows:  
## Final Answer  
#judge: <yes/no>  
#type: <vulnerability type>  
...
\end{lstlisting}

\textbf{Model:}  
\begin{lstlisting}[numbers=none,breakautoindent=false,breakindent=0pt]
Invoking `get_function_definition` with `{'func_name': 'ngx_pcalloc'}`  
\end{lstlisting}

\textbf{Tool:}  
\begin{lstlisting}[numbers=none,breakautoindent=false,breakindent=0pt]
[SUCCESS]: [File: src/core/ngx_palloc.c]  
void *
ngx_pcalloc(ngx_pool_t *pool, size_t size)
{
    void *p;

    p = ngx_palloc(pool, size);
    if (p) {
        ngx_memzero(p, size);
    }

    return p;
}
\end{lstlisting}

\textbf{Model:}  
\begin{lstlisting}[numbers=none,breakautoindent=false,breakindent=0pt]
Invoking `get_function_definition` with `{'func_name': 'ngx_array_init'}`  
\end{lstlisting}

\textbf{Tool:}  
\begin{lstlisting}[numbers=none,breakautoindent=false,breakindent=0pt]
[SUCCESS]: [File: src/core/ngx_array.h]  
static ngx_inline ngx_int_t
ngx_array_init(ngx_array_t *array, ngx_pool_t *pool, ngx_uint_t n, size_t size)
{
    array->nelts = 0;
    array->size = size;
    array->nalloc = n;
    array->pool = pool;

    array->elts = ngx_palloc(pool, n * size);
    if (array->elts == NULL) {
        return NGX_ERROR;
    }

    return NGX_OK;
}
\end{lstlisting}

\textbf{Model:}  
\begin{lstlisting}[numbers=none,
  breaklines=true,
  breakatwhitespace=true,
  breakindent=2em]
<reasoning>
We inspect ngx_core_module_create_conf:
1. Allocation and Initialization:
   ccf = ngx_pcalloc(cycle->pool, sizeof(ngx_core_conf_t)):
   - Calls ngx_palloc then zeroes size bytes. ngx_palloc fetches from the pool; if it fails, NULL is returned.
   - After ngx_pcalloc, ccf is either a valid zeroed block or NULL. The code immediately checks for NULL and returns, so no NULL-pointer dereference (CWE-476).
2. Field assignments:
   - The code writes to ccf->daemon, master, remote_admin, trace_enable, timer_resolution, shutdown_timeout, worker_processes, debug_points, rlimit_nofile, rlimit_core, user, group.
   - All writes are within the allocated struct; there is no indexing beyond bounds (no buffer overflow / CWE-119).
3. Array initialization:
   ngx_array_init(&ccf->env, cycle->pool, 1, sizeof(ngx_str_t)):
   - This sets nelts, size, nalloc, pool, then does array->elts = ngx_palloc(pool, 1 * sizeof(ngx_str_t)).
   - If allocation fails, ngx_array_init returns NGX_ERROR, and our code returns NULL without touching array->elts.
   - No out-of-bounds indexing or use-after-free; no pointer is dereferenced after being freed (CWE-416).
4. Pointer validity:
   - The only dereferences of cycle are cycle->pool in ngx_pcalloc and ngx_array_init.
   - The contract is that cycle must be a valid pointer passed by the caller (Nginx core).
   - There is no code path in this function that dereferences a NULL or freed pointer.
Conclusion: none of the listed CWEs apply.
</reasoning>
##Final Answer
#judge: no
#type: N/A
\end{lstlisting}

\end{tcolorbox}

\begin{figure*}[t]
  \centering
  \begin{minipage}[t]{0.32\textwidth}
    \centering
    \includegraphics[width=\linewidth]{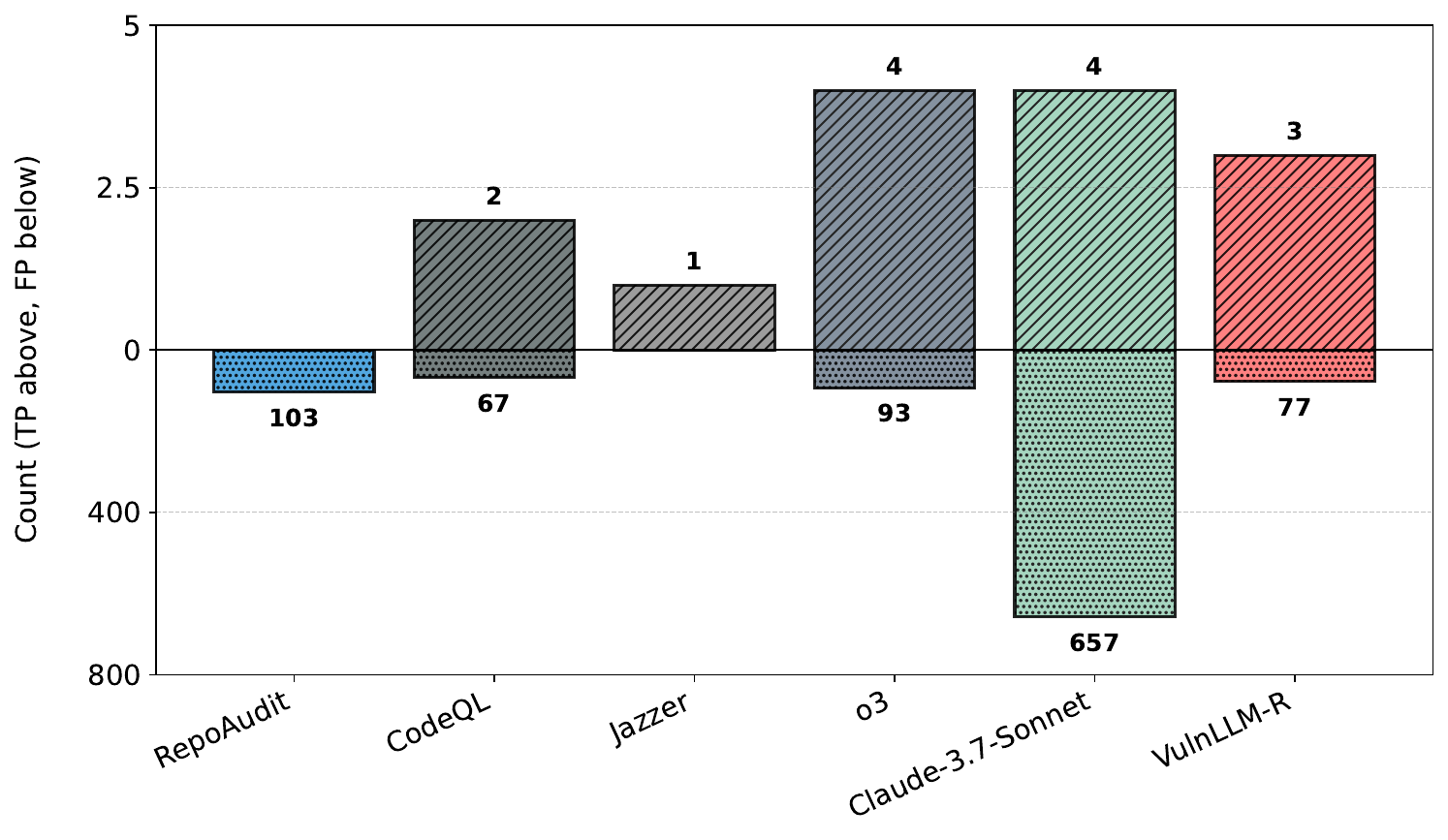}
    \subcap{(a) Tika (5 vulnerabilities, 2632 target functions)}
  \end{minipage}\hfill
  \begin{minipage}[t]{0.32\textwidth}
    \centering
    \includegraphics[width=\linewidth]{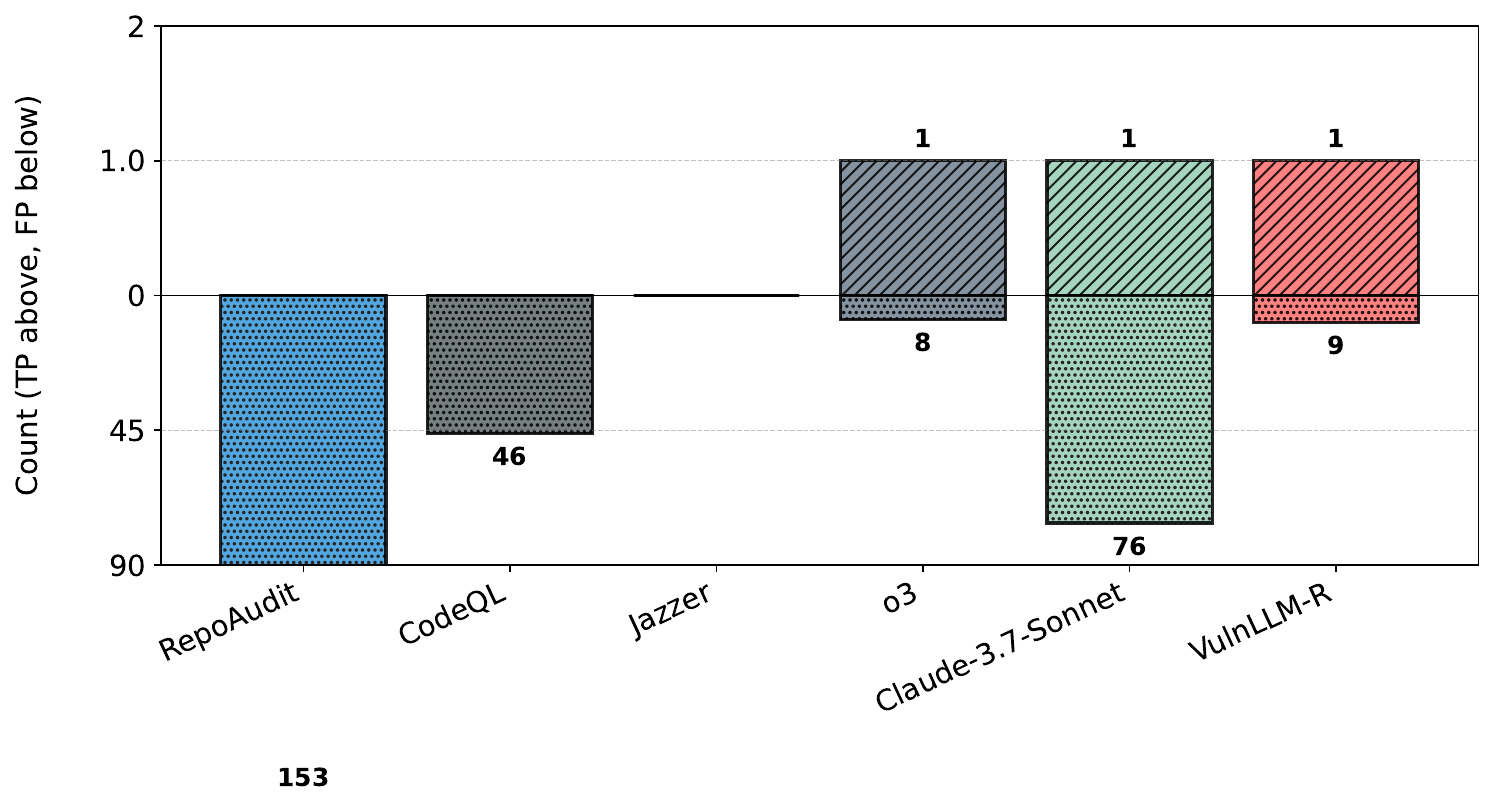}
    \subcap{(b) Zookeeper (1 vulnerability, 546 target functions)}
  \end{minipage}\hfill
  \begin{minipage}[t]{0.32\textwidth}
    \centering
    \includegraphics[width=\linewidth]{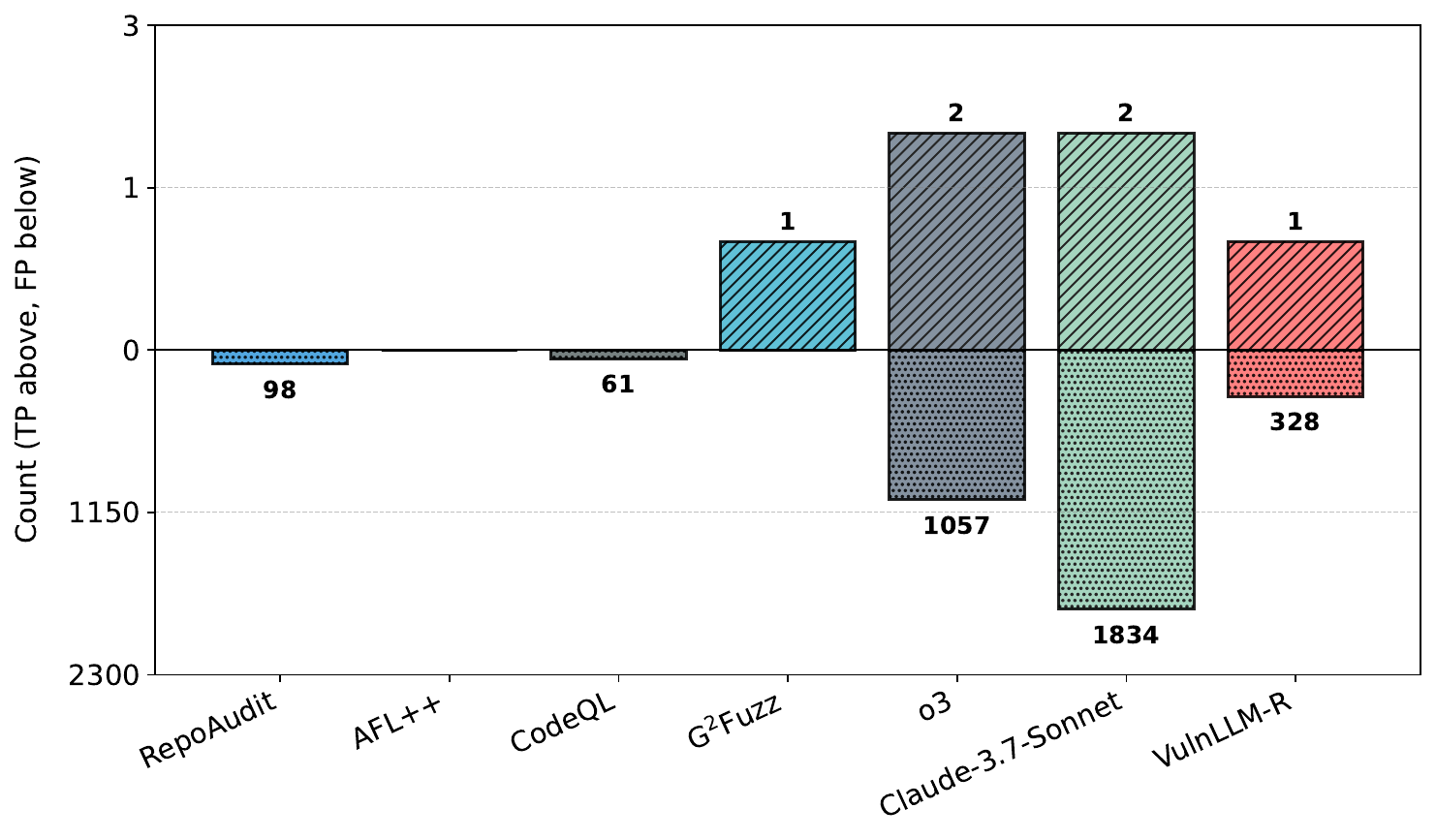}
    \subcap{(c) FreeRDP (2 vulnerabilities, 3347 target functions)}
  \end{minipage}

  \begin{minipage}[t]{0.48\textwidth}
    \centering
    \includegraphics[width=\linewidth]{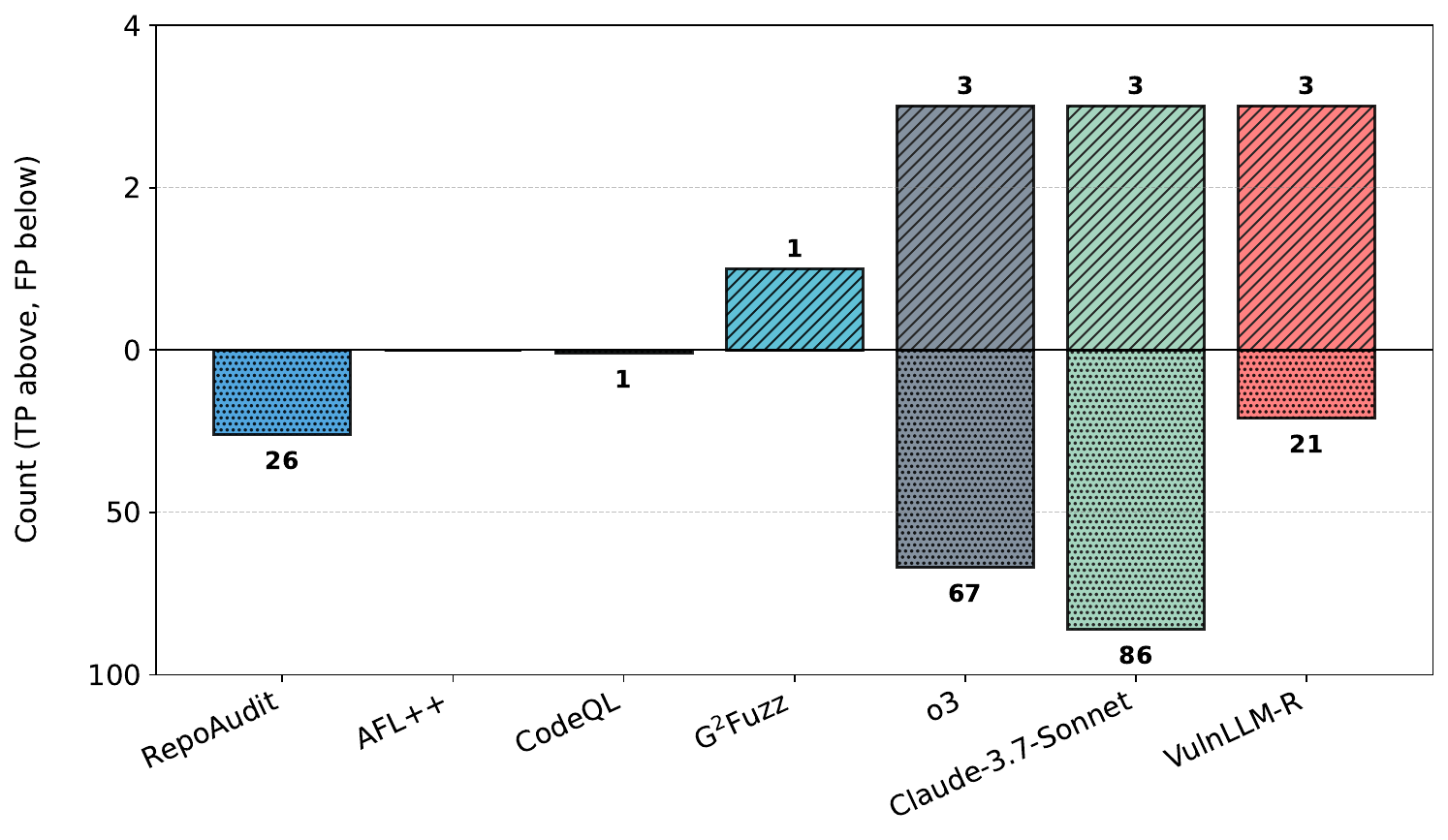}
    \subcap{(d) libexif (3 vulnerabilities, 225 target functions)}
  \end{minipage}\hfill
  \begin{minipage}[t]{0.48\textwidth}
    \centering
    \includegraphics[width=\linewidth]{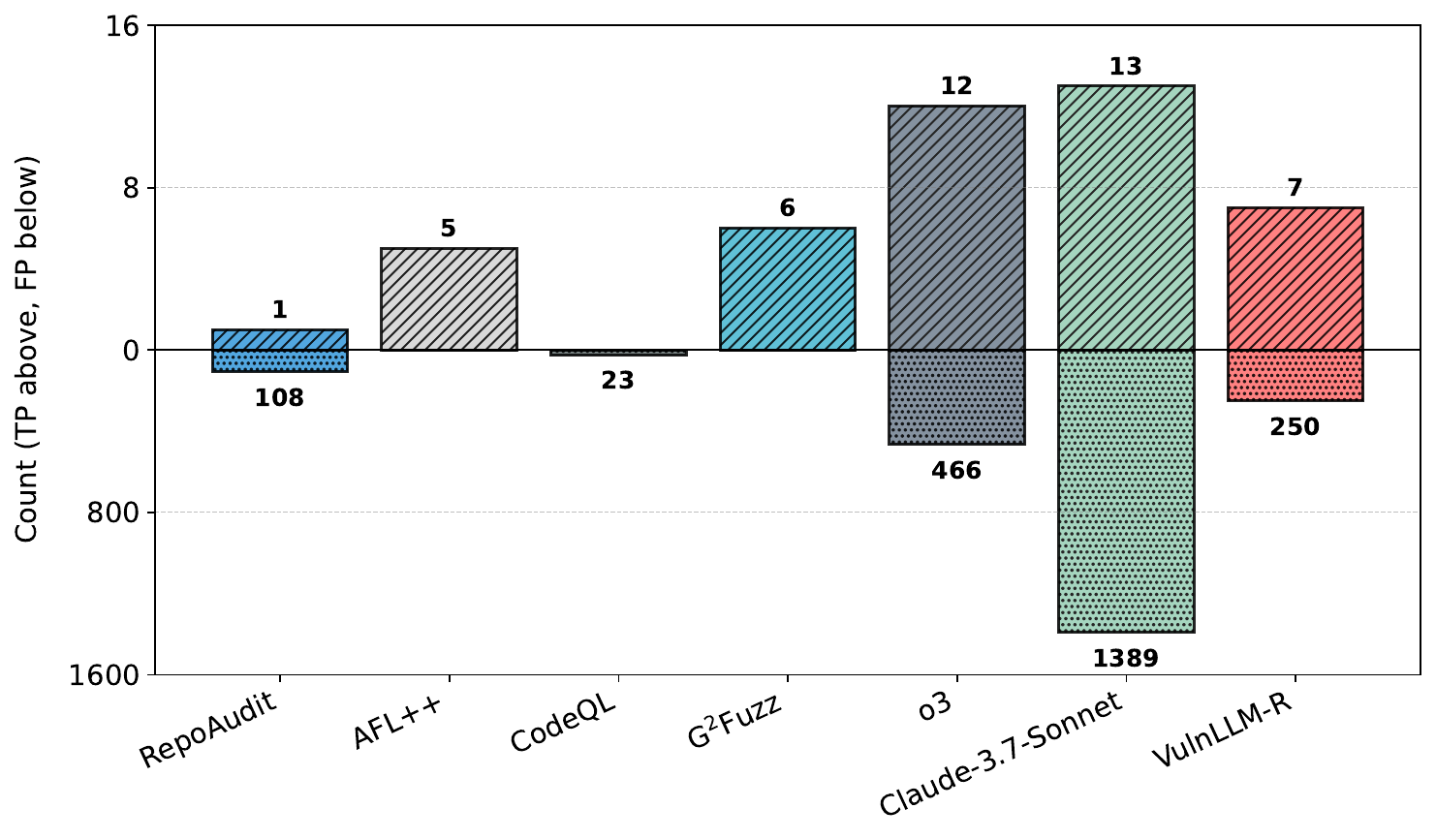}
    \subcap{(e) Nginx (14 vulnerabilities, 2076 target functions)}
  \end{minipage}

  \caption{Project-level TP/FP comparison of \sys{} and all baselines across five targets.}
  \label{fig:all_project_level_res}
\end{figure*}

\Cref{fig:all_project_level_res} shows the detailed project-level true positive and false positive results for each individual target, illustrating how \sys{} compares with all baselines across Tika, Zookeeper, FreeRDP, libexif, and Nginx.

\section{Templates of Prompt}
\label{app:prompt_template}
We provide the templates of the prompts used in our experiments.
First, we provide the template of the vulnerability detection prompt.
\begin{tcolorbox}[colback=white,title={Vulnerability Detection Prompt}, colframe=black, breakable]

You are an advanced vulnerability detection model. 
Your task is to check if a specific vulnerability exists in a given piece of code. 
The code may contain a long context, which is the stack trace of the function. 
They are separated by ``// context'' and ``// target function''. 
You need to output whether the target function is vulnerable and the type of vulnerability present with cwe id (CWE-xx). 

You are given the following code snippet:
\begin{verbatim}
{CODE}
\end{verbatim}

You should STRICTLY structure your response as follows:

\begin{verbatim}
## Final Answer
#judge: <yes/no>
#type: <vulnerability type>
\end{verbatim}

Additional Constraint:\\

If \texttt{\#judge: yes}, then \texttt{\#type:} must contain exactly one CWE.\\

If \texttt{\#judge: yes}, the model must output only the most probable CWE related to the given code snippet.\\

\textbf{Example}

If the target function is vulnerable to a CWE-79, you should finally output:
\begin{verbatim}
## Final Answer
#judge: yes
#type: CWE-79
\end{verbatim}

If the target function does not contain vulnerabilities related to the given CWE, you should finally output:
\begin{verbatim}
## Final Answer
#judge: no
#type: N/A
\end{verbatim}
    

\end{tcolorbox}

Second, we provide part of our human-written constitutions in~\Cref{tab:constitutions}.
\begin{table*}[htbp]
  \centering
  \caption{Part of our human-written constitutions}
  \label{tab:constitutions}
  \begin{tabular}{p{2cm}p{11cm}}
  \toprule
  \textbf{CWE ID} & \textbf{Constitution} \\
  \midrule
  CWE-134 & Ensure that format strings are fixed and not influenced by external input. \\
  CWE-191 & Look for arithmetic operations, especially decrement operations that are performed on variables that can potentially hold minimum integer values. \\
  CWE-22 & Confirm that the code includes checks for absolute paths, using security flags, and verify that the code correctly identifies and handles absolute paths by setting errors and returning failure codes when such paths are detected. \\
  CWE-327 & Identify the use of strong, well-regarded cryptographic algorithms, and understand that the use of such algorithms mitigates vulnerabilities by providing adequate cryptographic strength. \\
  CWE-367 & Identify that the benign code does not perform a separate check before using the resource, and understand that by directly attempting to use the resource, the code avoids the window of opportunity for a race condition. \\
  CWE-526 & Note the absence of conditional logic that prevents the exposure of environment variables. If the code always executes the output of an environment variable without any checks, it is likely vulnerable. \\
  CWE-121 & Ensure that both lower and upper bounds are checked before using an index to access an array, and recognize the addition of a condition of data size as a fix. \\
  CWE-23 & Recognize code that uses static, predefined strings for file paths, ensuring that no external input can influence the path construction. \\
  CWE-369 & Verify that the code includes checks to validate inputs before they are used in division operations. This includes ensuring the divisor is not zero or close to zero. \\
  CWE-400 & Ensure that input values are validated and constrained within safe limits before being used to control resource consumption. \\
  CWE-416 & Ensure that memory allocation and deallocation are handled correctly, with no operations on pointers after deallocation, and verify that any deallocated pointers are not used in subsequent operations. \\
  CWE-457 & Ensure that all elements of an array or structure are initialized before any use. This can be achieved by initializing the entire array in a loop before any other operations. \\
  CWE-476 & Ensure that pointers are validated before they are dereferenced. This includes checking if a pointer is NULL and handling such cases appropriately, often using conditional statements or error-handling constructs. \\
  CWE-758 & Identify code patterns where objects are used without proper initialization. Specifically, look for instances where a pointer is dereferenced to access or copy data from an uninitialized object. \\
  CWE-761 & Prefer index-based traversal over pointer arithmetic when iterating through a buffer. This ensures that the original pointer remains unchanged and can be safely freed. \\
  CWE-843 & Ensure that the type of data being accessed is consistent with the type of the variable it is pointing to. \\
  CWE-125 & Ensure that any operation involving buffers or arrays checks the boundaries before accessing elements. Look for conditions where the code might access elements beyond the allocated memory. \\
  CWE-190 & Ensure that input is validated before being used in arithmetic operations. This includes checking that the input is within a safe range to prevent overflow. \\
  CWE-787 & Identify code sections where data is written to buffers. Pay attention to calculations involving buffer sizes and offsets. Look for operations that modify buffer pointers or indices, especially in loops or conditional statements. \\
  \bottomrule
  \end{tabular}
\end{table*}

\section{More Results on Failed Attempts}
\label{app:comparedpo}

We explore DPO as an alternative training approach. However, as shown in~\Cref{tab:dpo_results}, \sys-DPO performs significantly worse than \sys across all languages, with F1 scores dropping to 0.601 for Python and 0.631 for C. The performance degradation is particularly pronounced on Java, the unseen language, where \sys-DPO achieves only 0.607 compared to \sys's 0.870. 
This suggests that DPO is less effective than our summary-based approach for vulnerability detection, especially when generalizing to new programming languages.

\begin{table}[ht]
  \centering
  \caption{DPO-based offline RL results for \sys across languages}
  \label{tab:dpo_results}
  \begin{tabular}{@{} l  c  c  c @{}}
    \toprule
    \multicolumn{1}{c}{\textbf{}} 
      & \multicolumn{1}{c}{\textbf{Python}} 
      & \multicolumn{1}{c}{\textbf{C}} 
      & \multicolumn{1}{c}{\textbf{Java}}\\
    \midrule
   \sys 
      & 0.723 & 0.737 & 0.870\\
    \sys-DPO
      & 0.601 & 0.631 & 0.607\\
    \sys-base
      & 0.675 & 0.570 & 0.690\\
    \bottomrule
  \end{tabular}

\end{table}

\end{document}